\documentstyle [12pt]{article}
\def\int {\intop \limits}

\def\fnote#1{\footnote}
\begin{document}
\newcommand{\dst}[1]{\displaystyle{#1}}
\newcommand{\barl}{\begin{array}{rl}}
\newcommand{\ball}{\begin{array}{ll}}
\newcommand{\ear}{\end{array}}
\newcommand{\barc}{\begin{array}{c}}
\newcommand{\sne}[1]{\displaystyle{\sum _{#1} }}
\newcommand{\sn}[1]{\displaystyle{\sum ^{\infty }_{#1} }}
\newcommand{\ini}[1]{\displaystyle{\int ^{\infty }_{#1}}}
\newcommand{\myi}[2]{\displaystyle{\int ^{#1}_{#2}}}
\newcommand{\inn}{\displaystyle{\int }}
\newcommand{\be}{\begin{equation}}
\newcommand{\ee}{\end{equation}}
\newcommand{\aq}[1]{\label{#1}}
\renewcommand \theequation{\thesection.\arabic{equation}}

\vspace*{4.0cm}
\centerline{\Large {\bf The theory of the Landau, Pomeranchuk,}}
\vskip .25cm
\centerline{\Large {\bf Migdal effect}}
\vskip .5cm
\centerline{\large{\bf V. N. Baier and V. M. Katkov}}
\centerline{Budker Institute of Nuclear Physics,
 630090 Novosibirsk, Russia}
\vskip 2.0cm
\begin{abstract}
Bremsstrahlung of photons from highly relativistic electrons
is investigated. The cross section of the processes, which is suppressed
due to a multiple scattering of an emitting electron in dense media
(LPM effect) and due to photon interaction with electrons of a medium,
is calculated with an accuracy up to "next to leading logarithm" and
with the Coulomb corrections taken into account. Making allowances for
a multiple scattering and a polarization of a medium an analysis of radiation
on a target boundary is carried out. The method of consideration of radiation
in a thin target under influence of the LPM effect is developed.
Interrelation with the recent experiment is discussed.
\end{abstract}

\newpage
\section{Introduction}

When a high-energy electron emits a soft photon via bremsstrahlung,
the process occurs over a rather long distance,
known as the formation length. If anything happens to an electron
or a photon while traveling this distance, the emission can be disrupted.
Landau and Pomeranchuk were the first who showed that if the formation
length of bremsstrahlung becomes comparable to the distance over which
a multiple scattering becomes important, the bremsstrahlung will be
suppressed \cite{1}. Migdal \cite{2}, \cite{3} developed a quantitative
theory of this phenomenon. Side by side with the multiple scattering of
emitting electron one has to take into account also an influence
of a medium on radiated electromagnetic field. Since long distances are
essential in the problem under consideration this can be done by
introducing dielectric constant $\varepsilon(\omega)$. This effect leads
also to suppression of the soft photon emission (Ter-Mikaelian
effect, see in \cite{4}).
A clear qualitative analysis of different mechanisms
of suppression is presented in
\cite{5},\cite{6}. More simple derivation of the Migdal's results is given
in \cite{7}.

The next step in a quantitative theory of LPM effect was made in
\cite{8}. This theory is based on the quasiclassical operator method
in QED developed by authors \cite{7}, \cite{9}. One of the basic equations
(obtained with use of kinetic equations describing a motion of
electron in a medium in the presence of external field) is
 the Schr\"{o}dinger equation in external field
with imaginary potential (Eq.(3.3),\cite{8}). The same equation
(without external field)
was rederived recently in \cite{10}. The last derivation is based on the
approach results of which is coincide basically with the
operator quasiclassical
method. In \cite{11} a new calculation approach is developed
where multiple scattering is described with the path integral treatment.

New activity with the theory of LPM effect is connected with a very
successful series of experiments \cite{12} - \cite{14} performed
at SLAC during last years (see in this connection \cite{15}).
In these experiments the cross section
of bremsstrahlung of soft photons with energy from 200~KeV to
500~MeV from electrons with energy 8~GeV and 25~GeV is measured
with an accuracy of the order of a few percent. Both LPM and dielectric
suppression is observed and investigated. These experiments are the
challenge for the theory since in all the mentioned papers calculations
are performed to logarithmic accuracy which is not enough for description
of the new experiment. The contribution of the Coulomb corrections (at least
for heavy elements) is larger then experimental errors and these corrections
should be taken into account.

In the present paper we calculated the cross
section of bremsstrahlung process with term $\propto 1/L$ , where $L$
is characteristic logarithm of the problem, and with the Coulomb corrections
taken into account (Section 2 and Appendix A). This cross section
is valid for very high energies when the LPM effect manifest itself for a
photon energy of the order of an energy of the initial electron.
In the photon energy region, where the LPM effect
is "turned off", our cross section
gives the exact Bethe-Heitler cross section (within power accuracy) with
Coulomb corrections. This important feature was absent in
the previous calculations. The polarization of a medium is incorporated into
this approach (Section 3). The considerable contribution into the soft part of
the investigated in the experiment spectrum of radiation gives a photon
emission on the boundaries of a target. We calculated this contribution
taking into account the multiple scattering and polarization of a medium
for the case when a target is much thicker than the
formation length of the radiation (Section 4). In Section 5 we considered
a case when a target is much thinner than the
formation length. In this case
the cross section has multiplicative form (probability of radiation
times cross section of scattering for the given impact parameter).
In Section 6 a case of an intermediate thickness of a target (between
cases of a thick and a thin target) is analyzed, polarization of
a medium is not included. In Section 7 a qualitative picture of
a spectral curve (an effective thickness of a target, position of a
minimum) is discussed. In Section 8 we compare the theoretical curve for the
intensity spectrum with the data.
Although agreement between experiment and theory is rather satisfactory,
an additional analysis should be done to obtain information about an
accuracy of agreement between experimental data and theory.

\section{The LPM effect in an infinitely thick target}

As well known (see, e.g. \cite{16}, Sec.93) the formation length of
radiation is (in this paper the system $\hbar=c=1$ is used)
\begin{equation}
l_c=\frac{2\varepsilon \varepsilon'}{m^2 \omega \zeta},\quad
\zeta=1+\gamma^2 \vartheta^2,\quad \varepsilon'=\varepsilon-\omega,\quad
\gamma=\frac{\varepsilon}{m},
\label{1}\end{equation}
where~ $\varepsilon$~is the energy of the initial electron,
~$\omega$~ is the energy of radiated photon,
$\vartheta$~ is the angle between momenta of the photon and
the initial electron. We consider first the case when the formation length
is much shorter than thickness of a target $l (l_c \ll l)$.
In this case the spectral distribution of the probability of radiation
per unit time is given by expression (2.18), \cite{8} (see also \cite{9},
Section 7.4)
\begin{equation}
\frac{dW}{d\omega}=\alpha \omega {\rm Re} \int_{0}^{\infty} d\tau
\exp (-i\frac{a\tau}{2})\left[\frac{\omega^2}{\gamma^2 \varepsilon'^2}
\varphi_0(0, \tau) -i \left(1+\frac{\varepsilon^2}{\varepsilon'^2} \right)
\mbox{\boldmath$\nabla$} \mbox{\boldmath$\varphi$}(0, \tau)\right],
\label{2}\end{equation}
where $\displaystyle{\alpha=e^2=\frac{1}{137}}$,
functions $\varphi_{\mu}(\varphi_0, \mbox{\boldmath$\varphi$})$
satisfy an equation
\begin{equation}
\frac{\partial \varphi_{\mu}}{\partial \tau}-\frac{ib}{2} \Delta
\varphi_{\mu} ({\bf x}, \tau)=n(\Sigma({\bf x})-\Sigma(0))
\varphi_{\mu} ({\bf x}, \tau)
\label{3}\end{equation}
with the initial conditions
\begin{equation}
\varphi_0({\bf x}, 0)=\delta({\bf x}),\quad
\mbox{\boldmath$\varphi$}({\bf x}, 0)=
-i \mbox{\boldmath$\nabla$} \delta({\bf x}).
\label{4}\end{equation}
Here $n$ is the number density of atoms in the medium,
${\bf x}$ is the coordinate in two-dimensional space conjugated
to the space (two-dimensional) of radiation angle
$\mbox{\boldmath$\vartheta$}$, $\Sigma({\bf x})$ is the Fourier
transform of the scattering cross section:
\begin{equation}
\displaystyle{\Sigma({\bf x})=\int_{}^{}d^2\vartheta
\exp (i{\bf x}\mbox{\boldmath$\vartheta$})
\sigma(\mbox{\boldmath$\vartheta$}),\quad
a=\frac{\omega m^2}{\varepsilon \varepsilon'},\quad
b=\frac{\omega \varepsilon}{\varepsilon'}}.
\label{5}\end{equation}
For a screened Coulomb potential we have
\begin{equation}
\displaystyle{\sigma(\vartheta)=
\frac{4Z^2\alpha^2}{\varepsilon^2(\vartheta^2+\vartheta_1^2)^2},\quad
\Sigma(x)=4\pi\frac{Z^2\alpha^2}{\varepsilon^2}
\frac{x}{\vartheta_1}K_1(x\vartheta_1)},
\label{6}\end{equation}
where $\displaystyle{\vartheta_1=\frac{1}{a_s\varepsilon}}$,
$a_s$ is the screening radius ($\displaystyle{a_s=0.81 a_B Z^{-1/3}, a_B}
$ - the Bohr radius),
$K_1$ is the modified Bessel function. As we will show below, the
main contribution to the probability is given by
\newline
$\displaystyle{\frac{1}{x}\sim \vartheta
\geq \frac{1}{\gamma} \gg \vartheta_1 =\frac{\lambda_c}{a_s\gamma}}$,
where $\displaystyle{\lambda_c=\frac{1}{m}=\left(\frac{\hbar}{mc}\right)}$
is the electron Compton wavelength. Expanding $K_1(x\vartheta_1)$ as a power
series in $x\vartheta_1$ and introducing new variables
\begin{equation}
t=\frac{a}{2}\tau,\quad \mbox{\boldmath$\varrho$} =
\sqrt{\frac{a}{b}}{\bf x} = \frac{1}{\gamma}{\bf x},
\label{7}\end{equation}
we obtain for the spectral distribution of the probability of radiation
\begin{equation}
\frac{dW}{d\omega}=\frac{2\alpha}{\gamma^2} {\rm Re} \int_{0}^{\infty} dt
e^{-it}\left[R_1
\varphi_0(0, t) + R_2
{\bf p} \mbox{\boldmath$\varphi$}(0, t)\right],
\label{8}\end{equation}
where $\displaystyle{R_1=\frac{\omega^2}{\varepsilon \varepsilon'},\quad
R_2 =\frac{\varepsilon}{\varepsilon'}+\frac{\varepsilon'}{\varepsilon}}$,
and the functions $\varphi_{\mu}$ now satisfy an equation
\begin{equation}
\begin{array}{ll}
&\displaystyle{i\frac{\partial \varphi_{\mu}}{\partial t}=\left({\bf p}^2
- iV(\mbox{\boldmath$\varrho$}) \right) \varphi_{\mu},\quad
{\bf p}= -i \mbox{\boldmath$\nabla$}_{\mbox{\boldmath$\varrho$}},\quad
V(\mbox{\boldmath$\varrho$})=-Q\mbox{\boldmath$\varrho$}^2
\Big(\ln \gamma^2 \vartheta_1^2}\\
&\displaystyle{+\ln \frac{\mbox{\boldmath$\varrho$}^2}{4}+2C-1\Big),\quad
Q=\frac{2\pi n Z^2 \alpha^2 \varepsilon \varepsilon'}{m^4 \omega},\quad
C=0.577...}
\end{array}
\label{9}\end{equation}
with the initial conditions $\varphi_0(\mbox{\boldmath$\varrho$}, 0)=
\delta(\mbox{\boldmath$\varrho$}),
\quad \mbox{\boldmath$\varphi$}(\mbox{\boldmath$\varrho$}, 0)
={\bf p} \delta(\mbox{\boldmath$\varrho$})$, the functions
$\varphi_0$ and $\mbox{\boldmath$\varphi$}$ in (\ref{8}) are rescaled
according with the initial conditions (factors $1/\gamma^2$ and
$1/\gamma^3$, correspondingly).
Note, that it is implied that in formulae (\ref{2}),(\ref{8}) subtraction
at $V=0$ is made.

The potential $V(\mbox{\boldmath$\varrho$})$
(\ref{9}) corresponds to consideration of scattering
in the Born approximation. The difference of exact as a function of
$Z\alpha$  potential $V(\mbox{\boldmath$\varrho$})$ and taken in
the Born approximation is computed in Appendix A.
The potential $V(\mbox{\boldmath$\varrho$})$ with the Coulomb corrections taken
into account is
\begin{equation}
\begin{array}{ll}
\displaystyle{V(\mbox{\boldmath$\varrho$})=-Q\mbox{\boldmath$\varrho$}^2
\Big(\ln \gamma^2 \vartheta_1^2
+\ln \frac{\mbox{\boldmath$\varrho$}^2}{4}+2C-1+2f(Z\alpha)\Big)}\\
\displaystyle{=-Q\mbox{\boldmath$\varrho$}^2
\Big(\ln \gamma^2 \vartheta_2^2
+\ln \frac{\mbox{\boldmath$\varrho$}^2}{4}+2C\Big)},
\end{array}
\label{9a}\end{equation}
where $\vartheta_2=\vartheta_1 \exp (f-1/2)$, the function $f=f(Z\alpha)$
see in (\ref{a10}).

In above formulae $\mbox{\boldmath$\varrho$}$ is space of the
impact parameters measured in the Compton wavelengths $\lambda_c$,
which is conjugate to space of the transverse momentum
transfers measured in the electron mass $m$.
An operator form of a solution of Eq. (\ref{9}) is
\begin{equation}
\begin{array}{ll}
&\varphi_0(\mbox{\boldmath$\varrho$}, t) =
\exp(-iHt) \varphi_0(\mbox{\boldmath$\varrho$}, 0)=
<\mbox{\boldmath$\varrho$}| \exp(-iHt) |0>,\quad
H={\bf p}^2-iV(\mbox{\boldmath$\varrho$}),\\
&\mbox{\boldmath$\varphi$}(\mbox{\boldmath$\varrho$}, t)=
\exp(-iHt) {\bf p} \varphi_0(\mbox{\boldmath$\varrho$},0)=
<\mbox{\boldmath$\varrho$}| \exp(-iHt){\bf p} |0>,
\end{array}
\label{10}\end{equation}
where we introduce the Dirac state vectors: $|\mbox{\boldmath$\varrho$}>$
is the state vector of coordinate $\mbox{\boldmath$\varrho$}$,
$<\mbox{\boldmath$\varrho$}|0>=\delta(\mbox{\boldmath$\varrho$})$.
Substituting (\ref{10}) into (\ref{8}) and taking integral over $t$
we obtain
for the spectral distribution of the probability of radiation
\begin{equation}
\frac{dW}{d\omega}=\frac{2\alpha}{\gamma^2} {\rm Im} <0|
R_1 \left(G^{-1}-G_0^{-1} \right)
+ R_2 {\bf p} \left(G^{-1}-G_0^{-1} \right) {\bf p} |0>,
\label{11}\end{equation}
where
\begin{equation}
G={\bf p}^2+1-iV,\quad G_0={\bf p}^2+1.
\label{12}\end{equation}
Here and below we consider an expression $<0|...|0>$ as a limit:
${\rm lim}~{\bf x} \rightarrow 0,
\newline {\rm lim}~{\bf x'} \rightarrow 0$ of
$<{\bf x}|...|{\bf x'}>$.

Now we estimate effective impact parameters $\varrho_c$ which give
the main contribution into radiation probability. Since characteristic
values of $\varrho_c$ will be found straightforwardly at calculation of
(\ref{11}), we estimate characteristic angles $\vartheta_c$ connected
with $\varrho_c$ by an equality $\varrho_c=1/(\gamma \vartheta_c)$.
The mean square scattering angle of a particle on the formation length of
a photon $l_c$ (\ref{1}) has the form
\begin{equation}
\vartheta_s^2=\frac{4\pi Z^2 \alpha^2}{\varepsilon^2}nl_c
\ln \frac{\zeta}{\gamma^2 \vartheta_1^2}=
\frac{4Q}{\gamma^2\zeta}
\ln \frac{\zeta}{\gamma^2 \vartheta_1^2}.
\label{13}\end{equation}
When $\vartheta_s^2 \ll 1/\gamma^2$ the contribution in the probability
of radiation gives a region where $\zeta \sim 1 (\vartheta_c=1/\gamma)$,
in this case $\varrho_c=1$. When $\vartheta_s \gg 1/\gamma$
the characteristic angle of radiation is determined by  self-consistency
arguments:
\begin{equation}
\vartheta_s^2 \simeq \vartheta_c^2 \simeq \frac{\zeta_c}{\gamma^2}=
\frac{4Q}{\zeta_c \gamma^2} \ln \frac{\zeta_c}{\gamma^2\vartheta_1^2},\quad
\frac{4Q}{\zeta_c^2} \ln \frac{\zeta_c}{\gamma^2\vartheta_1^2}=1,\quad
4Q\varrho_c^4 \ln \frac{1}{\gamma^2\vartheta_1^2 \varrho_c^2}=1.
\label{14}\end{equation}
It should be noted that when characteristic impact parameter
$\varrho_c$ becomes smaller than a radius of nucleus $R_n$,
the potential $V(\mbox{\boldmath$\varrho$})$ acquires an oscillator form
(see Appendix B, Eq.(\ref{b3}))
\begin{equation}
V(\mbox{\boldmath$\varrho$})=Q\mbox{\boldmath$\varrho$}^2
\left(\ln \frac{a_s^2}{R_n^2}-0.208 \right)
\label{15}\end{equation}

Allowing for estimates (\ref{14}) we present the potential
$V(\mbox{\boldmath$\varrho$})$ (\ref{9}) in the following form
\begin{equation}
\begin{array}{ll}
&\displaystyle{V(\mbox{\boldmath$\varrho$})=V_c(\mbox{\boldmath$\varrho$})+
v(\mbox{\boldmath$\varrho$}),\quad V_c(\mbox{\boldmath$\varrho$})=
q\mbox{\boldmath$\varrho$}^2,\quad q=QL,\quad
L=\ln \frac{1}{\gamma^2\vartheta_2^2\varrho_c^2}},\\
&\displaystyle{v(\mbox{\boldmath$\varrho$})=
-\frac{q\mbox{\boldmath$\varrho$}^2}{L}
\left(2C+\ln \frac{\mbox{\boldmath$\varrho$}^2}{4\varrho_c^2} \right)}.
\end{array}
\label{16}\end{equation}
The inclusion of the Coulomb corrections $f(Z\alpha)$ and -1 into
$\ln \vartheta_2^2$ diminishes effectively the correction
$v(\mbox{\boldmath$\varrho$})$ to the potential
$V_c(\mbox{\boldmath$\varrho$})$.
In accordance with such division of the potential we present propagators
in expression (\ref{11}) as
\begin{equation}
G^{-1}-G_0^{-1}=G^{-1}-G_c^{-1} + G_c^{-1}-G_0^{-1}
\label{17}\end{equation}
where
\[
G_c={\bf p}^2+1-iV_c,\quad G={\bf p}^2+1-iV_c-iv
\]
This representation of the propagator $G^{-1}$ permits one to expand it
over "perturbation" $v$. Indeed, with an increase of $q$ the relative
value of the perturbation is diminished
$\displaystyle{(\frac{v}{V_c} \sim \frac{1}{L})}$ since effective impact
parameters diminish and, correspondingly, the value of
logarithm $L$ in (\ref{16}) increases.
The maximal value of $L$ is determined by a size of a nucleus $R_n$
\begin{equation}
L_{max}=\ln \frac{a_{s2}^2}{R_n^2} \simeq 2
\ln \frac{a_{s2}^2}{\lambda_c^2}
\equiv 2 L_1,
\label{18}\end{equation}
where $a_{s2}=a_s \exp (-f+1/2)$. So, one can to redefine the parameters
$a_s$ and $\vartheta_1$ to include the Coulomb corrections.

The matrix elements of the operator $G_c^{-1}$ could be calculated
explicitly. The exponential parametrization of the propagator is
\begin{equation}
\displaystyle{G_c^{-1}=i\int_{0}^{\infty}dt e^{-it} \exp (-iH_ct),\quad
H_c={\bf p}^2 - iq \mbox{\boldmath$\varrho$}^2}
\label{19}\end{equation}
Below we will use matrix elements of the operator
$\displaystyle{\exp (-iH_ct)}$
\begin{equation}
<\mbox{\boldmath$\varrho$}_1|\exp(-iH_ct)|\mbox{\boldmath$\varrho$}_2>
\equiv K_c(\mbox{\boldmath$\varrho$}_1, \mbox{\boldmath$\varrho$}_2, t).
\label{20}\end{equation}
The function $K_c(\mbox{\boldmath$\varrho$}_1, \mbox{\boldmath$\varrho$}_2, t)$
satisfies the Schr\"odinger equation (\ref{9}) over each of two
(symmetrical) variables $\mbox{\boldmath$\varrho$}_1$ and
$\mbox{\boldmath$\varrho$}_2$ with
$V=q\mbox{\boldmath$\varrho$}^2$ and the initial condition
\begin{equation}
K_c(\mbox{\boldmath$\varrho$}_1, \mbox{\boldmath$\varrho$}_2, 0)=
\delta(\mbox{\boldmath$\varrho$}_2-\mbox{\boldmath$\varrho$}_1).
\label{21}\end{equation}
We will seek a solution in the form (see also \cite{8})
\[
K_c(\mbox{\boldmath$\varrho$}_1, \mbox{\boldmath$\varrho$}_2, t)=
\exp \left[\alpha(t)(\mbox{\boldmath$\varrho$}_1^2+
\mbox{\boldmath$\varrho$}_2^2)+
2\beta(t)\mbox{\boldmath$\varrho$}_1\mbox{\boldmath$\varrho$}_2+
\gamma(t) \right].
\]
Substituting this expression into (\ref{9}) we find a set of equations
for $\alpha, \beta, \gamma$
\begin{equation}
\dot{\alpha}=4i\alpha^2-q,\quad\dot{\beta}=4i\alpha \beta,\quad
\dot{\gamma}=4i\alpha.
\label{22}\end{equation}
The initial conditions for this set follows from definition (\ref{20}):
\begin{equation}
\begin{array}{ll}
\displaystyle{\lim_{t\to\ 0}
<\mbox{\boldmath$\varrho$}_1|\exp (-iH_ct)|\mbox{\boldmath$\varrho$}_2>
\rightarrow
<\mbox{\boldmath$\varrho$}_1|\exp (-iH_0t)|\mbox{\boldmath$\varrho$}_2>=}\\
\displaystyle{\frac{1}{(2\pi)^2}\int_{}^{}d^2p
\exp\left( i(\mbox{\boldmath$\varrho$}_2-\mbox{\boldmath$\varrho$}_1)
{\bf p}-i{\bf p}^2t \right)=\frac{1}{4\pi i t}
\exp \left(\frac{i\left(\mbox{\boldmath$\varrho$}_2-
\mbox{\boldmath$\varrho$}_1 \right)^2}{4t}\right)}\\
\displaystyle{\equiv
K_0(\mbox{\boldmath$\varrho$}_2,\mbox{\boldmath$\varrho$}_1,t)},
\end{array}
\label{20a}\end{equation}
where $H_0={\bf p}^2$. From (\ref{20a}) one has the initial conditions
at $t \rightarrow 0$
\begin{equation}
\alpha(t) \rightarrow \frac{i}{4t},\quad \beta(t) \rightarrow -\frac{i}{4\pi},
\quad \gamma(t) \rightarrow -\ln (4\pi it).
\label{20b}\end{equation}
The solution of the set (\ref{22}) satisfying these initial conditions is
\begin{equation}
\alpha(t)=\frac{i\nu}{4}\coth \nu t,\quad \beta(t)
= -\frac{i\nu}{4 \sinh \nu t},\quad \gamma(t)=-\ln (\sinh \nu t) +
\ln \frac{\nu}{4\pi i},
\label{23}\end{equation}
where $\displaystyle{\nu=2\sqrt{iq}}$.
As a result, we obtain the following expression for the sought function
\begin{equation}
K_c(\mbox{\boldmath$\varrho$}_1, \mbox{\boldmath$\varrho$}_2, t)=
\frac{\nu}{4\pi i \sinh \nu t} \exp \left\{ \frac{i\nu}{4}
\left[ (\mbox{\boldmath$\varrho$}_1^2+\mbox{\boldmath$\varrho$}_2^2)
\coth \nu t - \frac{2}{\sinh \nu t}
\mbox{\boldmath$\varrho$}_1\mbox{\boldmath$\varrho$}_2\right] \right\}.
\label{24}\end{equation}
Substituting formulae (\ref{19}) and (\ref{24}) in the expression
for the spectral distribution of the probability of radiation (\ref{11})
we have
\begin{equation}
\begin{array}{ll}
&\displaystyle{\frac{dW_c}{d\omega}=
\frac{\alpha}{2\pi \gamma^2} {\rm Im}~\Phi (\nu)},\\
&\displaystyle{\Phi(\nu)=\nu\int_{0}^{\infty} dt e^{-it}\left[R_1
\left(\frac{1}{\sinh z}-\frac{1}{z}\right)-i\nu R_2
\left( \frac{1}{\sinh^2z}- \frac{1}{z^2}\right) \right]},
\end{array}
\label{25}\end{equation}
where $z=\nu t$. This formula gives
the spectral distribution of the probability of radiation derived by
Migdal \cite{2}. However, here Coulomb corrections are
included into parameter $\nu$ in contrast to \cite{2}.

We now expand the expression $G^{-1}-G_c^{-1}$ over powers of $v$
\begin{equation}
G^{-1}-G_c^{-1}=G_c^{-1}(-iv)G_c^{-1}+G_c^{-1}(-iv)G_c^{-1}(-iv)G_c^{-1}
+...
\label{26}\end{equation}
Substituting this expansion in (\ref{17}) and then in (\ref{11})
we obtain decomposition of the probability of radiation.Let us note that
for $Q \ll 1$ the sum of the probability of radiation
$\displaystyle{\frac{dW_c}{d\omega}}$ (\ref{25}) and
the first term of the expansion (\ref{26}) gives the Bethe-Heitler spectrum
of radiation, see below (\ref{35}).
At $Q \geq 1$ the expansion
(\ref{26}) is a series over powers of $\displaystyle{\frac{1}{L}}$.
It is important that variation of the parameter $\varrho_c$ by a factor
order of 1 has an influence on the dropped terms
in (\ref{26}) only.

In accordance with (\ref{17}) and (\ref{26}) we present the
probability of radiation in the form
\begin{equation}
\frac{dW}{d\omega}=\frac{dW_c}{d\omega}+\frac{dW_1}{d\omega}+
\frac{dW_2}{d\omega}+...
\label{27}\end{equation}
The probability of radiation $\displaystyle{\frac{dW_c}{d\omega}}$ is defined by
Eq.(\ref{25}). In formula (\ref{11}) with allowance for (\ref{17})
there is expression
\begin{equation}
\begin{array}{ll}
\displaystyle{-i<0|G^{-1}-G_c^{-1}|0>=\int_{0}^{\infty}dt_1
\int_{0}^{\infty}dt_2 e^{-i(t_1+t_2)}\int_{}^{}d^2\varrho
K_c(0, \mbox{\boldmath$\varrho$}, t_1)
v(\mbox{\boldmath$\varrho$})K_c(\mbox{\boldmath$\varrho$}, 0, t_2)}\\
\displaystyle{+\int_{0}^{\infty}dt_1\int_{0}^{\infty}dt_2\int_{0}^{\infty}dt_3
e^{-i(t_1+t_2+t_3)}\int_{}^{}d^2\varrho_1\int_{}^{}d^2\varrho_2
K_c(0, \mbox{\boldmath$\varrho$}_1, t_1)v(\mbox{\boldmath$\varrho$}_1)
K_c(\mbox{\boldmath$\varrho$}_1, \mbox{\boldmath$\varrho$}_2, t_2)}\\
\displaystyle{\times v(\mbox{\boldmath$\varrho$}_2)
K_c(\mbox{\boldmath$\varrho$}_2, 0, t_3) + ...},
\end{array}
\label{28}\end{equation}
where the matrix element $K_c$ is defined by (\ref{24}).
The term $\displaystyle{\frac{dW_1}{d\omega}}$ in (\ref{27})
corresponds to the first term
(linear in $v$) in (\ref{28}). Substituting (\ref{24}) we have
\begin{equation}
\begin{array}{ll}
\displaystyle{\frac{dW_1}{d\omega}=\frac{2\alpha}{\gamma^2}
{\rm Re} \int_{0}^{\infty}dt_1\int_{0}^{\infty}dt_2
e^{-i(t_1+t_2)}\int_{}^{}d^2\varrho
v(\mbox{\boldmath$\varrho$})
\frac{q^2}{\pi^2\nu^2}\frac{1}{\sinh \nu t_1}\frac{1}{\sinh \nu t_2}}\\
\displaystyle{\times\exp\left[-\frac{q\varrho^2}{\nu}\left(\coth \nu t_1+
\coth \nu t_2 \right) \right]
\left[R_1+\frac{4q^2\varrho^2}{\nu^2\sinh \nu t_1\sinh \nu t_2}
R_2 \right]},
\end{array}
\label{29}\end{equation}
where $\displaystyle{\nu=2\sqrt{iq}}$. Substituting in (\ref{29}) the explicit
expression for $v(\mbox{\boldmath$\varrho$})$ and integrating over
$d^2\varrho$ and $d(t_1-t_2)$ we obtain
the following formula for the first correction
to the probability of radiation
\begin{equation}
\begin{array}{ll}
\displaystyle{\frac{dW_1}{d\omega}=-\frac{\alpha}{4\pi \gamma^2 L}
{\rm Im}~F(\nu);\quad F(\nu)= \int_{0}^{\infty}\frac{dz e^{-it}}{\sinh^2z}
\left[R_1f_1(z)-2iR_2f_2(z) \right]},\\
\displaystyle{f_1(z)=\left(\ln \varrho_c^2+\ln \frac{\nu}{i}
-\ln \sinh z-C\right)g(z) - 2\cosh z G(z)},\\
\displaystyle{f_2(z) = \frac{\nu}{\sinh z}
\left(f_1(z)-\frac{g(z)}{2} \right),\quad g(z)=z\cosh z - \sinh z},\\
\displaystyle{G(z)=\int_{0}^{z}\left(1-y\coth y\right)dy,\quad
t=t_1+t_2,~z=\nu t}
\end{array}
\label{30}\end{equation}

As it was said above (see (\ref{14}), (\ref{18})), $\varrho_c=1$ at
$|\nu^2|=\nu_1^2=4QL_1 \leq 1 (q=QL_1)$. If the parameter $\nu_1 > 1$,
the value of $\varrho_c$ is defined from the equation (\ref{14}),
where $\vartheta_1 \rightarrow \vartheta_2$, up to
$\varrho_c=R_n/\lambda_c$. Then one has
\begin{equation}
\displaystyle{\ln \varrho_c^2+\ln\frac{\nu}{i}=\frac{1}{2}
\ln (\varrho_c^4 4QL)-i\frac{\pi}{4}=-i\frac{\pi}{4}},\quad \varrho_c^4 4QL=1.
\label{31}\end{equation}
It follows from (\ref{31}) that expression (\ref{14}) for $\varrho_c^2$,
which we chose a priori, corresponds to the mean value of $\varrho^2$.
From the above analysis we have that the factor at $g(z)$ in expression
for $f_1(z)$ in (\ref{30}) can be written in the form
\begin{equation}
\displaystyle{(\ln \varrho_c^2+\ln\frac{\nu}{i}-\ln \sinh z-C) \rightarrow
(\ln \nu_0 \vartheta(1-\nu_0)-i\frac{\pi}{4}-\ln \sinh z -C)},
\label{32}\end{equation}
where
\begin{equation}
\nu_0^2 \equiv |\nu|^2=4q= 4 QL(\varrho_c)=
\frac{8\pi nZ^2\alpha^2 \varepsilon \varepsilon'}{m^4 \omega}L(\varrho_c),
\label{32a}\end{equation}
$\vartheta(x)$ is the Heaviside step function.

When a scattering is weak ($\nu_1 \ll 1$), the main contribution
in (\ref{30}) gives a region where $z \ll 1$. Then
\begin{equation}
\begin{array}{ll}
&\displaystyle{f_1(z) \simeq -(C + \ln (it))\frac{z^3}{3}+\frac{2}{9}z^3
=\frac{z^3}{3}(\frac{2}{3}-C-\ln (it))},\\
&\displaystyle{-{\rm Im}~F(\nu)=\frac{1}{9}{\rm Im}~\nu^2\left(R_2-
R_1 \right),\quad L \rightarrow L_1}.
\end{array}
\label{33}\end{equation}
The corresponding asymptotes of the function $\Phi(\nu)$ (\ref{25}) is
\begin{equation}
\displaystyle{\Phi(\nu) \simeq \frac{\nu^2}{6}\left(R_1+2R_2 \right),
\quad (|\nu| \ll 1)}
\label{34}\end{equation}
Combining the results obtained (\ref{33}) and (\ref{34}) we obtain
the spectral distribution of the probability of radiation in the case
when scattering is weak $(|\nu| \ll 1)$
\begin{equation}
\begin{array}{ll}
\displaystyle{\frac{dW}{d\omega}=\frac{dW_c}{d\omega}+\frac{dW_1}{d\omega}=
\frac{\alpha}{2\pi \gamma^2} {\rm Im}~\left[\Phi(\nu)-\frac{1}{2L}F(\nu)
\right]}\\
\displaystyle{=\frac{\alpha}{2\pi \gamma^2}\frac{2Q}{3}\left[R_1\left(
L_1-\frac{1}{3} \right)+2R_2\left(L_1+\frac{1}{6} \right) \right]}\\
\displaystyle{=\frac{4Z^2\alpha^3n}{3m^2\omega}\Bigg[\frac{\omega^2}
{\varepsilon^2}
\left(\ln \left(183Z^{-1/3} \right)-\frac{1}{6} -f(Z\alpha)\right)}\\
\displaystyle{+2\left(1+\frac{\varepsilon'^2}{\varepsilon^2} \right)
\left(\ln \left(183Z^{-1/3} \right)+\frac{1}{12}-f(Z\alpha) \right) \Bigg]},
\end{array}
\label{35}\end{equation}
where $L_1$ is defined in (\ref{18}).
This expression coincide with the known Bethe-Heitler formula for probability
of bremsstrahlung from high-energy electrons in the case of complete
screening (if one neglects the contribution
of atomic electrons) written down within power accuracy (omitted terms are of
the order of powers of $\displaystyle{\frac{1}{\gamma}}$) with the
Coulomb corrections, see e.g. Eq.(18.30)
in \cite{7}, or Eq.(3.83) in \cite{17}.

The integral in the function ${\rm Im}~F(\nu)$ (\ref{30}) which defines
the first correction to
the probability of radiation (\ref{30}) can be
transformed into the another form containing the real functions only
\begin{equation}
\begin{array}{ll}
\displaystyle{-{\rm Im}~F(\nu)=D_1(\nu_0)R_1+\frac{1}{s}D_2(\nu_0)R_2;\quad
s=\frac{1}{\sqrt{2}\nu_0}},\\
\displaystyle{D_1(\nu_0)=\int_{0}^{\infty}\frac{dz e^{-sz}}{\sinh^2z}
\left[d(z)\sin sz+\frac{\pi}{4}g(z)\cos sz \right],\quad
D_2(\nu_0)=\int_{0}^{\infty}\frac{dz e^{-sz}}{\sinh^3 z}}\\
\displaystyle{\times \left\{\left[d(z)-
\frac{1}{2}g(z)\right]\left(\sin sz+\cos sz\right)
+\frac{\pi}{4}g(z)\left(\cos sz - \sin sz\right)\right\}},\\\\
\displaystyle{d(z)=(\ln \nu_0 \vartheta(1-\nu_0)-\ln \sinh z -C)g(z)
-2\cosh z G(z)},
\end{array}
\label{36}\end{equation}
where the functions $g(z)$ and $G(z)$ are defined in (\ref{30}). The form
(\ref{36}) is convenient for numerical calculations. Note, that parameter
$s$ in (\ref{36}) is two times larger than used by Migdal \cite{2}.

At $\nu_0 \gg 1$ the function $F(\nu)$ (see (\ref{30}) and (\ref{32}))
can be written in the form
\begin{equation}
\displaystyle{F(\nu)= \int_{0}^{\infty}\frac{dz}{\sinh^2z}
\left[R_1f_1(z)-2iR_2f_2(z) \right]}.
\label{37}\end{equation}
Integrating over $z$ we obtain
\begin{equation}
\displaystyle{-{\rm Im}~F(\nu)=\frac{\pi}{4}R_1+\frac{\nu_0}{\sqrt{2}}
\left(\ln 2-C+\frac{\pi}{4} \right)R_2}.
\label{38}\end{equation}
Under the same conditions ($\nu_0 \gg 1$) the function ${\rm Im}~\Phi(\nu)$
(\ref{25}) is
\begin{equation}
\displaystyle{{\rm Im}~\Phi(\nu)=\frac{\pi}{4}R_1+\frac{\nu_0}{\sqrt{2}}R_2}.
\label{39}\end{equation}
Thus, at $\nu_0 \gg 1$ the relative contribution of the first correction
$\displaystyle{\frac{dW_1}{d\omega}}$ is defined by
\begin{equation}
\displaystyle{r=\frac{dW_1}{dW_c}=\frac{1}{2L(\varrho_c)}
\left(\ln 2-C+\frac{\pi}{4} \right) \simeq \frac{0.451}{L(\varrho_c)}},
\label{40}\end{equation}
where $\displaystyle{L(\varrho_c)=\ln \frac{a_{s2}^2}{\lambda_c^2 \varrho_c^2}}$.

In the above analysis we did not consider an inelastic scattering of a
projectile on atomic electrons. The potential
$V_e(\mbox{\boldmath$\varrho$})$ connected with this process
can be found from formula (\ref{9a}) by substitution
$Z^2 \rightarrow Z, \vartheta_1 \rightarrow \vartheta_e=0.153 \vartheta_1$
(an analysis of an inelastic scattering on atomic electrons as well as
the parameter $\vartheta_e$ can be found in \cite{17}).
The summary potential including both an elastic and an inelastic scattering
is
\begin{equation}
\begin{array}{ll}
\displaystyle{V(\mbox{\boldmath$\varrho$})+V_e(\mbox{\boldmath$\varrho$})
=-Q(1+\frac{1}{Z})\mbox{\boldmath$\varrho$}^2
\Big[\ln \gamma^2 \vartheta_2^2
+\ln \frac{\mbox{\boldmath$\varrho$}^2}{4}+2C +
\frac{1}{Z+1}\left(\ln \frac{\vartheta_e^2}{\vartheta_1^2}-2f\right)\Big]}\\
\displaystyle{=-Q_{ef}\mbox{\boldmath$\varrho$}^2
\Big(\ln \gamma^2 \vartheta_{ef}^2
+\ln \frac{\mbox{\boldmath$\varrho$}^2}{4}+2C \Big)},
\end{array}
\label{40a}\end{equation}
where
\[
\displaystyle{Q_{ef}=Q(1+\frac{1}{Z}),\quad \vartheta_{ef}
= \vartheta_1 \exp \left[\frac{1}{1+Z}\left(Zf(\alpha Z)-1.88
\right)-\frac{1}{2} \right]}.
\]

\section{An influence of the polarization of a medium}
\setcounter{equation}{0}

When one considers bremsstrahlung of enough soft photons $\omega \leq
\omega_0 \gamma$, one has to take into account the effect of a polarization
of the medium. In a dense medium the velocity of a photon propagation
differs from the light velocity in the vacuum since
the index of refraction $n(\omega) \neq 1$
\begin{equation}
n(\omega)=1-\frac{\omega_0^2}{2\omega^2},\quad \omega_0^2
=\frac{4\pi n \alpha}{m};\quad
1-\frac{k}{\omega} \simeq \frac{1}{2}\left(1-\frac{k^2}{\omega^2} \right)
=\frac{1}{2} \frac{\omega_0^2}{\omega^2}.
\label{41}\end{equation}
Because of this the formation length diminishes as well as
the probability of radiation (see \cite{4}, the
qualitative discussion may be found in \cite{6}).
For analysis we use the general expression for the probability
of radiation, see Eq.(2.1), \cite{8}. The factor in front of exponent
in this expression
(see Eq.(2.2), \cite{8}) contains two terms $A$ and ${\bf B}$, the term
$A$ is not changed and the term ${\bf B}$ contains combination
\begin{equation}
{\bf v}-\frac{{\bf k}}{\omega} \simeq  \mbox{\boldmath$\vartheta$}
+{\bf n}\frac{\kappa_0^2}{2\gamma^2},\quad
\kappa_0=\frac{\omega_0 \gamma}{\omega},
\label{42}\end{equation}
and its dependence on $\kappa_0$ (term of the order $1/\gamma^2$)
may be neglected also. With regard for the polarization of a medium
the formation length (\ref{1}) acquires a form
\begin{equation}
l_f = \frac{2\gamma^2}{\omega}\left[1+\gamma^2\vartheta^2 +
\left(\frac{\gamma \omega_0}{\omega}\right)^2 \right]^{-1}.
\label{42a}\end{equation}
So, the dependence on $\omega_0$ manifests itself in the exponent of Eq.(2.1),
\cite{8} and respectively in the exponent of (\ref{2}) only:
\begin{equation}
\displaystyle{a \rightarrow 2\frac{\omega \varepsilon}{\varepsilon-\omega}
\left(1-\frac{k}{\omega}v \right) \simeq a \kappa
\equiv \tilde{a}, \quad \kappa \equiv 1+\kappa_0^2}.
\label{43}\end{equation}
Performing the substitution $a \rightarrow \tilde{a}$
in formula (\ref{7}) we obtain for the potential (\ref{16})
\begin{equation}
\begin{array}{ll}
\displaystyle{V(\mbox{\boldmath$\varrho$}) \rightarrow
\tilde{V}(\tilde{\mbox{\boldmath$\varrho$}}) =\tilde{Q}
\tilde{\mbox{\boldmath$\varrho$}}^2
\left(L\left(\frac{\tilde{\varrho}}{2\sqrt{\kappa}}\right)-2C\right)
=\tilde{V}_c(\tilde{\mbox{\boldmath$\varrho$}})+
\tilde{v}(\tilde{\mbox{\boldmath$\varrho$}}),\quad \tilde{\varrho}=
|\tilde{\mbox{\boldmath$\varrho$}}|=\varrho \sqrt{\kappa}},\\
\displaystyle{ \tilde{V}_c(\tilde{\mbox{\boldmath$\varrho$}})=
\tilde{q}\tilde{\mbox{\boldmath$\varrho$}}^2,\quad \tilde{q}=
\tilde{Q}\tilde{L}(\tilde{\varrho_c}),\quad
\tilde{Q}=\frac{Q}{\kappa^2},\quad
\tilde{L}(\tilde{\varrho_c})=
\ln \frac{\kappa}{\gamma^2\vartheta_2^2\tilde{\varrho_c}^2}},\\
\displaystyle{\tilde{v}(\tilde{\mbox{\boldmath$\varrho$}})=
-\frac{\tilde{q}\tilde{\mbox{\boldmath$\varrho$}}^2}{\tilde{L}}
\left(2C+\ln \frac{\tilde{\mbox{\boldmath$\varrho$}}^2}
{4\tilde{\varrho_c}^2} \right)}.
\end{array}
\label{43a}\end{equation}
The substitution (\ref{43}) in the expression
for the probability of radiation (\ref{8}) gives
\begin{equation}
\displaystyle{R_1 \rightarrow R_1,\quad R_2 \rightarrow R_2 \kappa
\equiv \tilde{R}_2}.
\label{44}\end{equation}
The value of the parameter $\tilde{\varrho_c}$ in (\ref{43a}) is
determined by equation (compare with Eq.(\ref{31}))
\begin{equation}
4\tilde{\varrho_c}^4 \tilde{Q} \tilde{L}(\tilde{\varrho_c})=1,\quad
{\rm for} \quad 4\tilde{Q}\tilde{L}(1) \geq 1.
\label{44a}\end{equation}
In the opposite case $\tilde{\varrho_c}=1$ and 
this is possible in two intervals of the photon energy $\omega$:
\begin{enumerate}
\item for $\kappa_0 \ll 1$ when the multiple scattering and effects of
the polarization of a medium are weak;
\item for $\kappa_0 \gg 1$ when effects of
the polarization of a medium become stronger then effects of the multiple
scattering ($\nu_0 < \kappa$).
\end{enumerate}
In an intermediate region we substitute $\tilde{\varrho_c}^2 \rightarrow
\varrho_c^2 \kappa$ in Eq.(\ref{44a}). After it we obtain the equation for
$\varrho_c$ which coincides with Eq.(\ref{31}), see also (\ref{32a}):
\begin{equation}
\frac{1}{\varrho_c^4}=\nu_0^2(\varrho_c),\quad \nu_0^2(\varrho_c)=
4QL(\varrho_c).
\label{44b}\end{equation}
Thus, for $\tilde{\varrho_c} < 1$ we have
\begin{equation}
\tilde{\nu_0}=\sqrt{4\tilde{Q}\tilde{L}(\tilde{\varrho_c})}=
\frac{1}{\tilde{\varrho_c}^2}=\frac{1}{\varrho_c^2 \kappa}=
\frac{\nu_0}{\kappa},\quad \tilde{L}(\tilde{\varrho_c})=L(\varrho_c),
\label{44c}\end{equation}
while for $\tilde{\nu_0} < 1$ we have
\begin{equation}
\tilde{\nu_0}=\sqrt{4\tilde{Q}\tilde{L}(1)}=\frac{2}{\kappa}\sqrt{Q}
\ln \left(\frac{a_{s2}^2 \kappa}{\lambda_c^2} \right).
\label{44d}\end{equation}
The spectral distribution of the probability of radiation (\ref{35})
with allowance for polarization of a medium have the form
\begin{equation}
\displaystyle{\frac{dW}{d\omega}=
\frac{\alpha}{2\pi \gamma^2} {\rm Im}~
\left[\tilde{\Phi}(\tilde{\nu})-\frac{1}{2\tilde{L}(\tilde{\varrho}_c)}
\tilde{F}(\tilde{\nu}) \right]},
\label{45}\end{equation}
where
\[
\displaystyle{\tilde{\Phi}(R_1,R_2)=\Phi(R_1, \tilde{R}_2),\quad
\tilde{F}(R_1,R_2)=F(R_1, \tilde{R}_2)},
\]
We consider now the case when an influence the polarization of a medium
manifests
itself in the conditions of the strong LPM effect $(\nu_0 \gg 1)$.
This influence becomes essential for low energy photons, when
the mean square angle of the multiple scattering (\ref{14})
on the formation length of a photon becomes smaller than
$\displaystyle{\omega_0^2/\omega^2\left(\tilde{\nu}_0=\frac{\nu_0}{\kappa}
\leq 1,\quad \kappa_0^2 \gg 1\right)}$. Indeed, in the case
$\tilde{\nu}_0 \gg 1 (\nu_0 \gg \kappa_0^2)$ one can use
asymptotes of functions $\Phi(\nu)$ and $F(\nu)$ at $\nu_0 \gg 1$
(see (\ref{37}), (\ref{39})), we have
\begin{equation}
\begin{array}{ll}
\displaystyle{\frac{dW}{d\omega}=\frac{\alpha}{2\pi \gamma^2}R_2
\kappa \frac{\tilde{\nu}_0(1+\tilde{r})}{\sqrt{2}}
=\frac{\alpha}{2\pi \gamma^2}R_2 \frac{\nu_0}{\sqrt{2}}
(1+r)},\\
\displaystyle{\tilde{r} = \frac{0.451}{\tilde{L}(\tilde{\varrho}_c)},\quad
\tilde{L}(\tilde{\varrho}_c)=L(\varrho_c)=
\ln \frac{a_{s2}^2}{\lambda_c^2 \varrho_c^2}}.
\end{array}
\label{46}\end{equation}
In the opposite case $\nu_0 \ll \kappa_0^2$, the characteristic momentum
transfer in the used units ($\zeta_c$)
are defined by value $\kappa_0^2 (\tilde{\varrho}_c^2=1)$,
one can use asymptotic expansions (\ref{33}) and (\ref{34}) and we
have for the spectral distribution of the probability of radiation
\begin{equation}
\displaystyle{\frac{dW}{d\omega}=\frac{16}{3}\frac{Z^2\alpha^3n}
{m^2\omega \kappa_0^2} \left(L_{p}+\frac{1}{12}-f(Z\alpha) \right)=
\frac{4}{3\pi}\frac{Z^2\alpha^2\omega}{m\gamma^2}
\left(L_{p}+\frac{1}{12}-f(Z\alpha) \right)},
\label{47}\end{equation}
where $f(Z\alpha)$ is defined in (\ref{a10}),
$\displaystyle{L_{p}=\ln \left(183Z^{-1/3}\kappa_0 \right)}$.
The results obtained agree with given in \cite{4} where calculations
are fulfilled within logarithmic accuracy and without Coulomb corrections.
It is seen that a dependence of spectral distribution on photon
energy ($\omega d\omega$) differs essentially from the Bethe-Heitler one
($d\omega/\omega$), the probability is independent on density $n$.

The formula (\ref{47}) is applicable only up to value
$\kappa_0=\lambda_c/R_n$ or if $\omega > \omega_b$, where
\begin{equation}
\omega_b=\frac{R_n}{\lambda_c}\omega_0\gamma
\simeq \alpha Z^{1/3}\gamma \omega_0;\quad
\frac{\omega}{\varepsilon} > \alpha Z^{1/3}\frac{\omega_0}{m}.
\label{48}\end{equation}
For example, for electrons with energy $\varepsilon=25~GeV$ and
gold target ($\omega_0=80~eV$) one has $\omega_b \simeq 125~KeV$.
For $\omega < \omega_b$ one has take into account the form factor of
a nucleus (see Appendix B). In this case the argument of the logarithm
in (\ref{48}) ceases its dependence on photon energy $\omega$.
In the limit $\omega \ll \omega_b$
the spectral distribution of the probability of radiation
is
\begin{equation}
\displaystyle{\frac{dW}{d\omega}=
\frac{4}{3\pi}\frac{Z^2\alpha^2\omega}{m\gamma^2}
\left(\ln \frac{a_s}{R_n} - 0.02 \right)}
\label{49}\end{equation}

\section{A target of a finite thickness}
\setcounter{equation}{0}

In the case when a finiteness of a target is essential
the probability of radiation is defined not only by the relative
time $\tau=t_2-t_1$ as in Section 2.
The used radiation theory is formulated in terms of
two times (see eqs.(2.1) - (2.3) of \cite{8}).
Proceeding from this formulation we can obtain more general expression
which takes into account boundary effects. With allowance for
polarization of a medium we
have for the spectral distribution of the probability of radiation
\begin{equation}
\begin{array}{ll}
\displaystyle{\frac{dw}{d\omega}=\frac{4\alpha}{\omega}
{\rm Re}\int_{-\infty}^{\infty}dt_2
\int_{-\infty}^{t_2}dt_1 \exp \left(-i\mu(t_2)t_2+i\mu(t_1)t_1 \right)}\\
\displaystyle{\times \left[r_1 \varphi_0 (0, t_2, t_1)-ir_2
\mbox{\boldmath$\nabla$} \mbox{\boldmath$\varphi$}(0, t_2, t_1) \right]},
\end{array}
\label{50}\end{equation}
where
\begin{equation}
\begin{array}{ll}
\displaystyle{\mu(t)=\vartheta(-t)+\vartheta(T-t)+\kappa \vartheta(t)
\vartheta(T-t),\quad T=\frac{la}{2}=\frac{l\omega m^2}
{2\varepsilon \varepsilon'}},\\
\displaystyle{r_1 = \frac{\omega^2}{\varepsilon^2},\quad
r_2=1+\frac{\varepsilon'^2}{\varepsilon^2},\quad \kappa=1+\kappa_0^2},
\end{array}
\label{51}\end{equation}
here $l$ is the thickness of a target,
$\kappa_0$ is defined in (\ref{42}). So, we split time interval
(in the used units) into three parts: before target ($t<0$), after target
($t>T$) and inside target ($0 \leq t \leq T$). The functions
$\varphi_{\mu}(\mbox{\boldmath$\varrho$}, t_2, t_1),\quad
\varphi_{\mu}=\varphi_{\mu}(\varphi_0, \mbox{\boldmath$\varphi$})$ satisfy
the equation (\ref{9}), but now the potential $V$ depends on time
\begin{equation}
\begin{array}{ll}
\displaystyle{i\frac{\partial \varphi_{\mu}}{\partial t}
={\cal H}(t) \varphi_{\mu},\quad {\cal H} (t)={\bf p}^2-
iV(\mbox{\boldmath$\varrho$})g(t),\quad g(t)=\vartheta(t)\vartheta(T-t)};\\
\displaystyle{\varphi_0(\mbox{\boldmath$\varrho$},t_1,t_1)=
\delta(\mbox{\boldmath$\varrho$}),\quad
\mbox{\boldmath$\varphi$}(\mbox{\boldmath$\varrho$},t_1,t_1)={\bf p}
\delta(\mbox{\boldmath$\varrho$})}.
\end{array}
\label{52}\end{equation}
Using an operator
form of a solution of Eq. (\ref{52}) (compare with (\ref{10})) we can
present the probability (\ref{50}) in the form
\begin{equation}
\begin{array}{ll}
\displaystyle{\frac{dw}{d\omega}= \frac{4\alpha}{\omega}{\rm Re}
\int_{-\infty}^{\infty}dt_2
\int_{-\infty}^{t_2}dt_1 \exp \left(-i\mu(t_2)t_2+i\mu(t_1)t_1 \right)}\\
\displaystyle{\times \left<0|r_1 S(t_2,t_1)+r_2 {\bf p}
S(t_2, t_1) {\bf p}|0 \right>,\quad S(t_2, t_1)={\rm T} \exp
\left[-i\int_{t_1}^{t_2} {\cal H}(t) dt \right]},
\end{array}
\label{53}\end{equation}
where the symbol ${\rm T}$ means the chronological product. Note, that
in (\ref{50}) and (\ref{53}) it is implied that subtraction is made at
$V=0$,~$\mu(t)=1~(\kappa=1)$.

Integrals over time in (\ref{53}) we present as integrals over
four domains:
\begin{enumerate}
\item $t_1 \leq 0,~0 \leq t_2 \leq T$;
\item $ 0 \leq t_1 \leq T,~0 \leq t_2 \leq T$;
\item $0 \leq t_1 \leq T,~ t_2 \geq T$;
\item $t_1 \leq 0,  t_2 \geq T$;
\end{enumerate}
in two more domains $t_{1,2} \leq 0$ and
$t_{1,2} \geq T$ an electron is moving entirely free and there is no
radiation. We consider in this Section the case, when
the thickness of a target $L$ is much larger than formation length
$l_f$ (\ref{42a}) or $(\nu_0+\kappa)T \gg 1$.
In this case domain 4) doesn't contribute.
The contributions of other domains are
\begin{equation}
\begin{array}{ll}
\displaystyle{I_1 \simeq \int_{-\infty}^{0}dt_1 \int_{0}^{\infty}dt_2
\exp \left(i(t_1-\kappa t_2) \right) \exp \left(-iHt_2 \right)
\exp \left(i H_0 t_1 \right)=-\frac{1}{H+\kappa}
\frac{1}{H_0+1}},\\
\displaystyle{I_2=\int_{0}^{T} dt_2 \int_{0}^{t_2} dt_1 \exp\left(-i
(H+\kappa)(t_2-t_1) \right) \simeq T \int_{0}^{\infty} d\tau
\exp \left(-i(H+\kappa)\tau \right)}\\
\displaystyle{- \int_{0}^{\infty} \tau d\tau
\exp \left(-i(H+\kappa)\tau \right)
=-i\frac{T}{H+\kappa}+\frac{1}{(H+\kappa)^2},\quad
I_3 \simeq - \frac{1}{ H_0+1}\frac{1}{H+\kappa}},
\end{array}
\label{54}\end{equation}
where $ H_0= {\bf p}^2$.
The term in $I_2$:~$-iT/(H+\kappa)$  describes the
probability of radiation considered in previous Sections. All other terms
define the probability of radiation of boundary photons \footnote[1]
{Radiation of boundary photons in an inhomogeneous electromagnetic field
was considered in \cite{18}.}. So, making mentioned subtraction we have
for the spectral distribution of the probability of radiation
of boundary photons
\begin{equation}
\begin{array}{ll}
\displaystyle{\frac{dw_b}{d\omega}=\frac{4\alpha}{\omega}
{\rm Re}\left<0|r_1 M+ r_2 {\bf p} M {\bf p}|0 \right>,\quad
M=M_V^{(1)}+M_V^{(2)}+M_0};\\
\displaystyle{M_V^{(1)}=\frac{2}{({\bf p}^2+\kappa)({\bf p}^2+1)}
-\frac{2}{(H+\kappa)({\bf p}^2+1)},~
M_V^{(2)}=\frac{1}{(H+\kappa)^2}-\frac{1}{({\bf p}^2+\kappa)^2}},\\
\displaystyle{M_0=\frac{1}{({\bf p}^2+1)^2}-\frac{2}{({\bf p}^2
+\kappa)({\bf p}^2+1)}+\frac{1}{({\bf p}^2+\kappa)^2}},\\
\end{array}
\label{55}\end{equation}
For a convenience here we made the subtraction
in two stages: first \newline (in $M_V^{(1)},~M_V^{(2)}$) we subtracted
terms with $V=0$ and second (in $M_0$) we subtracted terms with
both $V=0$ and $\kappa=1$.

We consider important case when both the LPM effect and the polarization of
a medium are essential. We will
calculate the main term with $V(\mbox{\boldmath$\varrho$})
=V_c(\mbox{\boldmath$\varrho$})$, see (\ref{16}).
Needed combinations are
\begin{equation}
\begin{array}{ll}
\displaystyle{\left<0\left|\frac{1}{                                                 H+\kappa}\frac{1}{{\bf p}^2+1}
\right|0 \right>}\\
\displaystyle{=-\int_{0}^{\infty}dt_1\int_{0}^{\infty}dt_2
\exp \left(-i(t_1+\kappa t_2) \right)\int_{}^{}d^2\varrho
K_c(0, \mbox{\boldmath$\varrho$}, t_2)
K_0(\mbox{\boldmath$\varrho$}, 0, t_1)},\\
\displaystyle{\left<0\left|\frac{1}{(H+\kappa)^2}\right|0 \right>=
-\int_{0}^{\infty}tdt \exp (-i\kappa t)K_c(0, 0, t),
\quad \left<0|M_0|0\right>= \frac{1}{4\pi}},\\
\displaystyle{\left<0|{\bf p}M_0{\bf p}|0\right>=\frac{\pi}{(2\pi)^2}
\int_{0}^{\infty}dp^2 p^2 M_0 = \frac{1}{4\pi}
\left[ \left( 1+\frac{2}{\kappa-1}\right)\ln \kappa -2 \right]},
\end{array}
\label{56}\end{equation}
where the functions $K_0(\mbox{\boldmath$\varrho$}_2,
\mbox{\boldmath$\varrho$}_1,t)$ and $K_c(\mbox{\boldmath$\varrho$}_2,
\mbox{\boldmath$\varrho$}_1,t)$ are defined in (\ref{20a}) and (\ref{24}).
Substituting into (\ref{56}) the explicit expressions
for these functions, calculating the vector derivatives as indicated
in (\ref{55}) we have for contribution of the first
term in (\ref{56})
\begin{equation}
\begin{array}{ll}
\displaystyle{\frac{dw_b^{(1)}}{d\omega}=-\frac{2\alpha}{\pi \omega}r_2~
{\rm Re}~\nu^2 \int_{0}^{\infty} dt_1 \int_{0}^{\infty} dt_2
\exp \left(-i(t_1+\kappa t_2) \right)
\bigg[\frac{1}{\left(\sinh \nu t_2 + \nu t_1 \cosh \nu t_2 \right)^2}}\\
\displaystyle{-\frac{1}{\left(\nu t_1 +\nu t_2 \right)^2} \bigg]}\\
\displaystyle{=-\frac{2\alpha}{\pi \omega}r_2~{\rm Im}~
\nu \int_{0}^{\infty} dt_1 \int_{0}^{\infty} dt_2
\exp \left(-i(t_1+t_2) \right)
\left[\frac{1}{\tanh \tilde{\nu}t_2+\nu t_1}-
\frac{1}{\tilde{\nu}t_2+\nu t_1} \right]},
\end{array}
\label{57}\end{equation}
where $\tilde{\nu}=\nu/\kappa$, the second term in the square brackets
is the subtraction term in accordance with (\ref{55}) (the
term $M_V^{(1)}$). For practical use it is convenient to write the
probability (\ref{57}) using real variables.
After some transformations it can be written  as
\begin{equation}
\begin{array}{ll}
\displaystyle{\frac{dw_b^{(1)}}{d\omega}=\frac{2\alpha}{\pi \omega}r_2
\int_{0}^{\infty}dt \exp (-t)\left(\cos t +\sin t \right)
\int_{0}^{t} dy \Bigg[\frac{1}{t-y+s \tanh \left(y/\kappa s
\right)}}\\
\displaystyle{-\frac{1}{t-y+y/\kappa} \Bigg]},
\end{array}
\label{59}\end{equation}
where $\displaystyle{s=\frac{1}{\sqrt{2}\nu_0}}$,
parameter $\nu_0$ is defined in
(\ref{32a}). Repeating the same operations with the second term in (\ref{56})
(this is the contribution of the term $M_V^{(2)}$ in (\ref{55})) we have
\begin{equation}
\begin{array}{ll}
\displaystyle{\frac{dw_b^{(2)}}{d\omega}=\frac{\alpha}{\pi \omega}r_2~
{\rm Re}~\nu^2 \int_{0}^{\infty} tdt
\exp \left(-i\kappa t\right)
\left[\frac{1}{\sinh^2\nu t}-\frac{1}{(\nu t)^2} \right]}\\
\displaystyle{=\frac{\alpha}{\pi \omega}r_2~ {\rm Re}~
\int_{0}^{\infty} dz
\exp \left(-i\frac{z}{\tilde{\nu}}\right)
\left[\frac{z}{\sinh^2 z}-\frac{1}{z} \right]}\\
\displaystyle{=\frac{\alpha}{\pi \omega}r_2
\int_{0}^{\infty}dz \exp \left(-\tilde{s}z \right)
\cos \tilde{s}z
\left[\frac{z}{\sinh^2 z}-\frac{1}{z} \right]},
\end{array}
\label{60}\end{equation}
where $\displaystyle{z=\nu t,~\tilde{s}=\frac{1}{\sqrt{2}\tilde{\nu}_0},~
\tilde{\nu}_0=\nu_0/\kappa}$.
The contribution of the term $M_0$ in (\ref{55})
is calculated in (\ref{56})
\begin{equation}
\displaystyle{\frac{dw_b^{(3)}}{d\omega}=
\frac{\alpha}{\pi \omega}\left\{r_1+r_2 \left[ \left( 1+
\frac{2}{\kappa-1}\right)\ln \kappa -2 \right]\right\}}.
\label{61}\end{equation}
The complete expression
for the spectral distribution of the probability of radiation
of boundary photons, in the case when both the LPM effect and the polarization of
a medium are taken into account, is
\begin{equation}
\displaystyle{\frac{dw_b}{d\omega}=\sum_{k=1}^{3} \frac{dw_b^{(k)}}{d\omega}}.
\label{62}\end{equation}

We consider now the limiting case when LPM effect is very strong
($\tilde{\nu}_0 \gg 1$). In this case we find for probabilities
in formulae (\ref{59}) and (\ref{60})
\begin{equation}
\begin{array}{ll}
\displaystyle{\frac{dw_b^{(1)}}{d\omega}=\frac{2\alpha}{\pi \omega}r_2
\left[\ln \tilde{\nu}_0-C-\frac{\ln \kappa}{\kappa-1}+
\frac{\pi^2}{8\sqrt{2}\tilde{\nu}_0}+\frac{1}{\sqrt{2}\tilde{\nu}_0}
\left(\ln \tilde{\nu}_0+1-C+\frac{\pi}{4} \right) \right]},\\
\displaystyle{\frac{dw_b^{(2)}}{d\omega}=\frac{\alpha}{\pi \omega}r_2
\left[1-\ln 2\tilde{\nu}_0 +C-\frac{\pi^2}{6\sqrt{2}\tilde{\nu}_0} \right]}.
\end{array}
\label{63}\end{equation}
Substituting asymptotes obtained and (\ref{61}) into (\ref{62}) we have
\begin{equation}
\displaystyle{\frac{dw_b}{d\omega}=
\frac{\alpha}{\pi \omega}\left\{r_1+r_2 \left[\ln \nu_0 -1 -C -
\ln 2+\frac{\sqrt{2}}{\nu_0}\left(\kappa\frac{\pi^2}{24}+
\ln \nu_0 +1-C\frac{\pi}{4} \right) \right]\right\}}.
\label{64}\end{equation}
As one can expect, the probability of radiation at $\nu_0 \gg 1+\kappa_0^2$
depends on the polarization of a medium in the term $\propto 1/\nu_0$ only.

In the opposite case $\tilde{\nu}_0 \ll 1 (\nu_0 \gg 1)$, the probabilities
$dw_b^{(1)},~dw_b^{(2)} \propto \tilde{\nu}_0^4$ and probability of
radiation of boundary photons is determined by the polarization of a medium.
Just in this case radiation of boundary photons is known as the
transition radiation:
\begin{equation}
\displaystyle{\frac{dw_b}{d\omega} \simeq \frac{dw_b^{(3)}}{d\omega}=
\frac{\alpha}{\pi \omega}\left\{r_1+r_2 \left[ \left( 1+
\frac{2}{\kappa-1}\right)\ln \kappa -2 \right]\right\}}.
\label{65}\end{equation}

In the case of weak LPM effect $\nu_0 \ll 1~(\omega \ll \varepsilon)$ we have
\begin{equation}
\displaystyle{\frac{dw_b}{d\omega} \simeq \frac{\alpha}{\pi \omega} r_2
\left(-\frac{2}{21}\nu_0^4 \right)}.
\label{65a}\end{equation}
In this case what we calculated as the boundary photons contribution
is actually correction (very small) to the probability
$\displaystyle{l\frac{dW}{d\omega}}$ (\ref{35}) which in this case has
additional (suppression) factor $\displaystyle{1-\frac{16}{21}\nu_0^4}$
which follows from the decomposition of the function {\rm Im}~$\Phi$.

The LPM effect for the case of structured targets (with many boundaries)
was analyzed recently in \cite{16a}.
The radiation of the boundary photons with
regard for the multiple scattering was considered in \cite{17a}
(for $\omega \ll \varepsilon$),
the polarization of a medium  was added in \cite{18a} and \cite{18b}.
Our results, which are consistent with obtained \cite{18a}, are presented
in more convenient for application form and the Coulomb corrections
are included.
In these papers the probability of radiation of boundary photons
(under condition of applicability of Eq.(\ref{64})) was analyzed also
to within the logarithmic accuracy (see Eq.(20) in \cite{18a} and
Eq.(15 in \cite{18b})). This accuracy is insufficient for parameters
connected with experiment \cite{12}-\cite{14}. For example, for
$\varepsilon=25~GeV$ and heavy elements the value $\nu_0$ equates
$\kappa$ for $\nu_0 \sim 20$. One can see from Eq.(\ref{64}) that in this case
$\ln \nu_0$ is nearly completely compensated by constant terms.

\section{A thin target}
\setcounter{equation}{0}

Finally we consider a situation when the formation length of
radiation is much larger than the thickness $l$ of a target
(a thin target, $l_c \gg l$). In this case the radiated photon
is propagating in the vacuum and one can neglect
the polarization of a medium.

Operator $S(t_1, t_2)$ (\ref{53}) we present in the form
\begin{equation}
\begin{array}{ll}
\displaystyle{S(t_2, t_1)=T \exp
\left[-i\int_{t_1}^{t_2} {\cal H}(t) dt \right]=
\exp \left(-i H_0 t_2 \right) {\cal L}(t_2, t_1)
\exp \left(iH_0 t_1 \right)};\\
\displaystyle{{\cal L}(t_2, t_1)=\exp \left(iH_0 t_2 \right) S(t_2, t_1)
\exp \left(-iH_0 t_1 \right)}.
\end{array}
\label{66}\end{equation}
Differentiating the operator ${\cal L}(t_2, t_1)$ over the first of
arguments we obtain
\begin{equation}
\frac{\partial {\cal L}(t, t_1)}{\partial t}=
-\exp \left(iH_0 t \right)V(\mbox{\boldmath$\varrho$},t) S(t, t_1)
\exp \left(-iH_0 t_1 \right)= -V(\mbox{\boldmath$\varrho$} +
2{\bf p}t, t) {\cal L}(t, t_1),
\label{67}\end{equation}
where $V(\mbox{\boldmath$\varrho$}, t)=V(\mbox{\boldmath$\varrho$}) g(t)$
(see (\ref{9}), (\ref{52})).
The formal solution of this equation with the initial condition
${\cal L}(t_1, t_1)=1$ has the form
\begin{equation}
{\cal L}(t_2, t_1)={\rm T} \exp
\left[-\int_{t_1}^{t_2}  dt V(\mbox{\boldmath$\varrho$} +
2{\bf p}t, t)\right],
\label{68}\end{equation}
where ${\rm T}$ means the chronological product.
This solution is exact. Now we take into account that we
are considering a short characteristic time contributing into
integral (\ref{68}), or more precisely
\begin{equation}
t \leq T =\frac{l a}{2},\quad l \ll l_c =\frac{2}{a \zeta},\quad
T \ll \frac{1}{\zeta},
\label{69}\end{equation}
where $l_c,~\zeta$ are defined in (\ref{1}). Since the main contribution
give $p \sim \sqrt{\zeta},
\newline \varrho \sim 1/\sqrt{\zeta},~
pt \ll 1/\sqrt{\zeta} \sim \varrho$~,where $p$ is characteristic mean
value of operator $|{\bf p}|$, one can neglect by the term
$2{\bf p}t$ in (\ref{68}), so that
\begin{equation}
{\cal L}(t_2, t_1) \simeq \exp
\left[-\int_{t_1}^{t_2}  dt V(\mbox{\boldmath$\varrho$}, t)\right].
\label{70}\end{equation}
In the probability of radiation enters the expression (cp~(\ref{11}),
(\ref{53}))
\begin{equation}
\begin{array}{ll}
\displaystyle{\left<0 \left|\exp \left(-iH_0t_2 \right)\left({\cal L}-1
\right) \exp \left(iH_0t_1 \right)\right|0 \right>}\\
\displaystyle{=\int_{}^{}d^2\varrho \left({\cal L}-1 \right)
\left< 0 \left| \exp \left(-iH_0t_2 \right)
\right| \mbox{\boldmath$\varrho$} \right> \left< \mbox{\boldmath$\varrho$}
\left| \exp \left(iH_0t_1 \right) \right| 0 \right>}.
\end{array}
\label{71}\end{equation}
Using an explicit form (\ref{20a}) of the matrix element
$\left< 0 \left| \exp \left(-iH_0t_2 \right)
\right| \mbox{\boldmath$\varrho$} \right>$
and neglecting terms of the order $\sim T (l/l_c)$ one
obtains starting from (\ref{53})
for the spectral distribution of the probability of radiation
\begin{equation}
\begin{array}{ll}
\displaystyle{\frac{dw_{th}}{d\omega}=\frac{\alpha}{4\pi^2 \omega}
\int_{-\infty}^{0} \frac{dt_1}{t_1} \int_{0}^{\infty} \frac{dt_2}{t_2}
\int_{}^{}d^2\varrho \left(r_1+r_2{\bf p}_1{\bf p}_2 \right)}\\
\displaystyle{\times \exp \bigg[-i\left(t_2 - t_1 \right)
+i\frac{\mbox{\boldmath$\varrho$}^2}{4}\left(\frac{1}{t_2}
-\frac{1}{t_1} \right) \bigg]\left(\exp (-VT) -1 \right)}\\
\displaystyle{=\frac{\alpha}{\pi^2 \omega} \int_{}^{} d^2\varrho
\left[r_1 K_0^2(\varrho)+r_2 K_1^2(\varrho) \right]
\left(1-\exp (-VT) \right)},
\end{array}
\label{72}\end{equation}
where ${\bf p}_1~({\bf p}_2)$ is the operator
${\bf p}=-i \mbox{\boldmath$\nabla$}$ acting on the function of
$\mbox{\boldmath$\varrho$}^2/t_1~(\mbox{\boldmath$\varrho$}^2/t_2)$,
$K_n$ is the modified Bessel function. Here we took into account that
in our case contribute domain $|t_1|,|t_2| \gg T$ and $t_1 \leq 0,
t_2 \geq 0$ since in domains $t_{1,2} \leq 0$ and
$t_{1,2} \geq T$ an electron is moving entirely free and there is no
radiation.
In implicit form the factorization contained in (\ref{72}) is presented in
\cite{18c}.
If
\newline $V(\varrho=1)T \ll 1$ one can
expand the exponent (the contribution of the region $\varrho \gg 1$
is exponentially damped because in this region
$K_{0,1}(\varrho) \propto \exp (-\varrho)$). In the first order
over $VT$ using the explicit expression for the potential
(\ref{9a}) we have to calculate following integrals:
\begin{equation}
\begin{array}{ll}
\displaystyle{\int_{0}^{\infty} K_0^2(\varrho)\varrho^3 d\varrho =\frac{1}{3},
\quad \int_{0}^{\infty} K_0^2(\varrho)\ln \varrho \varrho^3 d\varrho
=\frac{1}{3} \left(\ln 2 - C +\frac{1}{6} \right)};\\
\displaystyle{\int_{0}^{\infty} K_1^2(\varrho)\varrho^3 d\varrho =\frac{2}{3},
\quad \int_{0}^{\infty} K_1^2(\varrho)\ln \varrho \varrho^3 d\varrho
=\frac{2}{3} \left(\ln 2 - C -\frac{1}{12} \right)}
\end{array}
\label{72a}\end{equation}
Substituting these integrals  one obtains in this case
the Bethe-Heitler formula with the Coulomb corrections (\ref{35}).

We analyze now the opposite case when the multiple scattering
of a particle traversing a target is strong ($V(\varrho=1)T \gg 1$, the
mean square of multiple scattering angle $\vartheta_s^2 \gg 1/\gamma^2$).
We present the function $V(\varrho)T$ (see (\ref{9}), (\ref{9a}) and
(\ref{18})) as
\begin{equation}
\begin{array}{ll}
\displaystyle{V(\varrho)T= \frac{\pi Z^2 \alpha^2 n l}{m^2}\varrho^2
\left(\ln \frac{4a_{s2}^2}{\lambda_c^2 \varrho^2}-2C \right)=
A\varrho^2 \ln \frac{\chi_t}{\varrho^2}=
A\varrho^2 \left( \ln \frac{\chi_t}{\varrho_t^2} -
\ln \frac{\varrho^2}{\varrho_t^2} \right)},\\
\displaystyle{=k\varrho^2\left(1-\frac{1}{L_t}\ln \frac{\varrho^2}
{\varrho_t^2}\right);\quad A\varrho_t^2 \ln \frac{\chi_t}{\varrho_t^2}=1,\quad
L_t=\ln \frac{\chi_t}{\varrho_t^2} \simeq
\ln \frac{4a_{s2}^2}{\lambda_c^2 \varrho_t^2}},
\end{array}
\label{73}\end{equation}
where $\varrho_t$ is the lower boundary of values contributing into the
integral over $\varrho$.
Substituting this expression into (\ref{72}) we have the integral
\begin{equation}
\displaystyle{2\pi\int_{0}^{\infty}\varrho d\varrho K_1^2(\varrho)
\left\{1-\exp \left[-k\varrho^2 \left(1-\frac{1}{L_t}
\ln \frac{\varrho^2}{\varrho_t^2} \right) \right] \right\} \equiv \pi J}.
\label{74}\end{equation}
In this integral we expand the exponent in the integrand over $1/L_t$ keeping
the first term of the expansion. We find
\begin{equation}
\begin{array}{ll}
\displaystyle{J=J_1+J_2,\quad J_1=2\int_{0}^{\infty} K_1^2(\varrho)
\left[1-\exp \left(-k\varrho^2\right)\right] \varrho d\varrho}\\
\displaystyle{=2k\int_{0}^{\infty}d\varrho \varrho^3\left[K_0(\varrho)
K_2(\varrho)-K_1^2(\varrho) \right] \exp \left(-k\varrho^2\right)},\\
\displaystyle{J_2=-\frac{2k}{L_t}\int_{0}^{\infty} K_1^2(\varrho)
\exp \left(-k\varrho^2\right)\ln \frac{\varrho^2}{\varrho_t^2}
\varrho^3 d\varrho}
\end{array}
\label{75}\end{equation}
In the integral $J_1$ we performed an integration by parts. In the integrals
in (\ref{75}) it is convenient to substitute $z=k\varrho^2$ then
\begin{equation}
\begin{array}{ll}
\displaystyle{J_1=\frac{1}{k}\int_{0}^{\infty}
\left[K_0\left(\sqrt{\frac{z}{k}}\right)K_2\left(\sqrt{\frac{z}{k}}\right)
-K_1^2\left(\sqrt{\frac{z}{k}}\right) \right]\exp(-z) zdz},\\
\displaystyle{J_2=-\frac{1}{kL_t}\int_{0}^{\infty}
K_1^2\left(\sqrt{\frac{z}{k}}\right) \exp(-z) \ln z zdz}.
\end{array}
\label{76}\end{equation}
Expanding the modified Bessel functions $K_n(x)$ at $x \ll 1$ and taking the
integrals in the last expression
we have
\begin{equation}
\begin{array}{ll}
\displaystyle{J=J_1+J_2=\left(1+\frac{1}{2k}\right)\left(\ln 4k -C\right)+
\frac{1}{2k}-1 +\frac{C}{L_t}},\\
\displaystyle{k=\frac{\pi Z^2\alpha^2}{m^2}
nl\left(L_t+1-2C \right)}.
\end{array}
\label{77}\end{equation}
In the term with $K_0^2$ in (\ref{72}) the region $\varrho \sim 1$
contributes. So we have
\begin{equation}
\displaystyle{J_3=2 \int_{0}^{\infty} K_0^2(\varrho)
\left(1-\exp (-VT) \right)\varrho d\varrho \simeq
2 \int_{0}^{\infty} K_0^2(\varrho)
\varrho d\varrho=1}.
\label{78}\end{equation}
Substituting found $J$ and $J_3$ into (\ref{72}) we
obtain for the spectral distribution of the probability of radiation
in a thin target at conditions of the strong multiple scattering
\begin{equation}
\frac{dw_{th}}{d\omega}=\frac{\alpha}{\pi \omega}
\left(r_1+r_2 J\right).
\label{79}\end{equation}
The logarithmic term in this formula is well known in theory of the
collinear photons radiation at scattering of a radiating particle on angle
much larger than characteristic angles of radiation $\sim 1/\gamma$.
It is described with logarithmic accuracy in a quasi-real electron
approximation (see \cite{20}, Appendix B2).

The formula (\ref{72}) presents the probability of radiation in the case
when the formation length $l_c \gg l$. It is known,see e.g. \cite{7}, that
in this case a process of scattering of a particle is independent of
a radiation process and a differential probability of radiation at
scattering with the momentum transfer
${\bf q}$ can be presented in the form
\begin{equation}
dW_{\gamma}=dw_s({\bf q})dw_r({\bf q}, {\bf k}),
\label{80}\end{equation}
where $dw_s({\bf q})$ is the differential probability
of scattering with the momentum transfer ${\bf q}$ which depends on
properties of a target. The function $dw_r({\bf q}, {\bf k})$
is the probability of radiation of a photon with a momentum ${\bf k}$
when an emitting electron acquires the momentum transfer ${\bf q}$.
This probability has a universal form which is independent of properties
of a target. For an electron traversing an amorphous medium this
fact is reflected in formula (\ref{72}). Indeed, passing on to a
momentum space we have
\begin{equation}
\begin{array}{ll}
\displaystyle{dw_r({\bf q}, {\bf k})=
\frac{\alpha d\omega}{\pi^2 \omega} \int_{}^{} d^2\varrho
\left[r_1 K_0^2(\varrho)+r_2 K_1^2(\varrho) \right]
\left(1-\exp (-i{\bf q}\mbox{\boldmath$\varrho$}) \right)}\\
\displaystyle{=\frac{\alpha d\omega}{\pi \omega}\left[r_1 F_1\left
(\frac{q}{2}\right)+r_2 F_2\left(\frac{q}{2}\right) \right]};\\
\displaystyle{F_1(x)=1-\frac{\ln \left(x+\sqrt{1+x^2} \right)}{x\sqrt{1+x^2}},
\quad F_2(x)=\frac{2x^2+1}{x\sqrt{1+x^2}}\ln \left(x+\sqrt{1+x^2} \right)-1}.
\end{array}
\label{81}\end{equation}
Remind that $q$ is measured in electron mass. The probability of radiation in
this form was found in \cite{18a}.
For a differential probability of scattering (here we consider
the multiple scattering) there is a known formula (cp~(\ref{5}),
(\ref{6}) and (\ref{9}))
\begin{equation}
\begin{array}{ll}
\displaystyle{dw_s({\bf q})=F_s({\bf q})d^2q,\quad F_s({\bf q})=
\frac{1}{(2\pi)^2}\int_{}^{}d^2\varrho
\exp \left(-i{\bf q} \mbox{\boldmath$\varrho$}\right)
\exp \left(-V_s(\varrho) l\right)},\\
\displaystyle{V_s(\varrho)=
n\int_{}^{}d^2q\left(1-\exp(-i{\bf q}\mbox{\boldmath$\varrho$}) \right)
\sigma({\bf q})},
\end{array}
\label{82}\end{equation}
where $\sigma({\bf q})$ is the cross section of single scattering.

Using the formula (\ref{81}) one can easily obtain to within
logarithmic accuracy expressions (\ref{79}),(\ref{64}).
Both a radiation of boundary photons and a radiation in a thin
target may be considered as a radiation of collinear photons
(see e.g. \cite{20}) in the case when an emitting particle deviates 
at large angle ($\vartheta_s \gg 1/\gamma, q \gg 1$).
Using (\ref{81}) at $x \gg 1$ we find
\begin{equation}
\begin{array}{ll}
\displaystyle{dw_r(q) \simeq \frac{\alpha d\omega}{\pi \omega}\left[
r_1+r_2\left(\ln q^2 -1 \right) \right]};\\
\displaystyle{\int_{}^{}d^2q dw_r(q) F_s({\bf q})
\simeq \frac{\alpha d\omega}{\pi \omega}
\left[r_1+r_2\left(\ln \overline{q^2} -1 \right) \right]}.
\end{array}
\label{83}\end{equation}
For a thin target value of $\overline{q^2}$ is defined by mean
square of multiple scattering angle on a thickness of a target $l$,
and for boundary photons is the same but
on the formation length $l_f$. However,
if we one intends to perform computation beyond a logarithmic accuracy,
the method given in this Section has advantage since there is no necessity
to calculate $F_s({\bf q})$ and in our approach a problem of
calculation of the Coulomb corrections is solved in a rather simple way.

\section{A target of an intermediate thickness $l \sim l_c$}
\setcounter{equation}{0}

It appears that used in Section 4 approach
permits one to consider an important case when $l \sim l_c$.

According to the partition of integrals over time in formula (\ref{53})
into four domains we can write the probability of radiation as
\begin{equation}
\displaystyle{\frac{dw}{d\omega}=\sum_{n=1}^{4} \frac{dw_n}{d\omega},\quad
\frac{dw_n}{d\omega}=\frac{4\alpha}{\omega}{\rm Re}
\left(r_1 I_n^{(1)}+r_2 I_n^{(2)} \right)}.
\label{84}\end{equation}
The integrals in $I_n^{(1,2)}$ we compute on the assumption:
$\nu_0 \gg 1,~T \ll 1,~\nu_0 T \sim 1,~\kappa=1 $.
Since integrals in $I_n^{(1)}$
don't contain the logarithmic divergence, only the domain 4 contributes.
In the domains 1-3 one of the integrals in $I_n^{(1)}$ contains an
integration over an interval $0 \leq t \leq T$ and due to this reason
$dw_{1,2,3} \propto T \ll 1$. So, we consider $I_4^{(1)}$
\begin{equation}
\begin{array}{ll}
\displaystyle{I_4^{(1)}=\int_{-\infty}^{0}dt_1 \int_{T}^{\infty} dt_2
\exp (-i(t_1+t_2))\big<0\big|\exp (-i(H_0(t_2-T))) \exp (-iHT)}\\
\displaystyle{\times \exp(iH_0t_1)
-\exp(iH_0(t_2-t_1))\big| 0\big>
=\int_{0}^{\infty}dt_1\int_{0}^{\infty}dt_2 \exp(-i(t_1+t_2+T))}\\
\displaystyle{\times \int_{}^{}d^2\varrho_1\int_{}^{} d^2\varrho_2
K_0(0, \mbox{\boldmath$\varrho$}_1, t_1)\left[
K_c(\mbox{\boldmath$\varrho$}_1, \mbox{\boldmath$\varrho$}_2, T)
-K_0(\mbox{\boldmath$\varrho$}_1, \mbox{\boldmath$\varrho$}_2, T) \right]
K_0(\mbox{\boldmath$\varrho$}_2, 0, t_2)}.
\end{array}
\label{85}\end{equation}
Here a calculation of integrals over $\mbox{\boldmath$\varrho$}_1$ and
$\mbox{\boldmath$\varrho$}_2$ may be performed e.g. in a such way:
\begin{itemize}
\item an integral over relative angle between
$\mbox{\boldmath$\varrho$}_1$ and $\mbox{\boldmath$\varrho$}_2$
gives $J_0(\beta \varrho_1 \varrho_2)$ where $J_0(x)$ is the Bessel
function, $\displaystyle{\beta=\beta_c=\frac{\nu}{2\sinh \nu T}}$ and
$\displaystyle{\beta=\beta_0=\frac{1}{2 T}}$ for the first and second terms
in the square brackets in the right-hand side of (\ref{85}),
\item the remaining integrals over $\varrho_1$ and $\varrho_2$
can be found in tables.
\end{itemize}
So, we have
\begin{equation}
\begin{array}{ll}
\displaystyle{I_4^{(1)}=\frac{1}{4\pi i}\int_{0}^{\infty}dt_1
\int_{0}^{\infty}dt_2 \exp \left(-i(t_1+t_2+T) \right)
\left[N(t_1, t_2)-\frac{1}{t_1+t_2+T} \right]},\\
\displaystyle{N(t_1, t_2)=\frac{\nu}{(1+\nu^2t_1t_2)\sinh \nu T +
\nu (t_1+t_2)\cosh \nu T}}.
\end{array}
\label{86}\end{equation}
For $\nu_0 \gg 1$ the contribution into integral the term with $N(t_1,t_2)$
is of the order of $1/\nu_0$ and this term may be neglected. With allowance
for $T \ll 1$ we find
\begin{equation}
\begin{array}{ll}
\displaystyle{I_4^{(1)}=-\frac{1}{4\pi i}\int_{0}^{\infty}dt_1
\int_{0}^{\infty}dt_2 \exp \left(-i(t_1+t_2) \right) \frac{1}{t_1+t_2}}\\
\displaystyle{=-\frac{1}{8\pi i}\int_{0}^{\infty}\frac{dx}{x}
{\rm e}^{-ix}\int_{-x}^{x}dy
=-\frac{1}{4\pi i}\int_{0}^{\infty} {\rm e}^{-ix} = \frac{1}{4\pi}},
\end{array}
\label{87}\end{equation}
where $x=t_1+t_2,~y=t_1-t_2$.

The contribution of the domain 4 into the term with $r_2$
($I_4^{(2)}$ in (\ref{84})) contain two additional operators ${\bf p}$
(see (\ref{53})) which result additional factor
$\displaystyle{-\frac{\mbox{\boldmath$\varrho$}_1
\mbox{\boldmath$\varrho$}_2}
{4t_1 t_2}}$ in the integrand. Integration over the relative angle between
$\mbox{\boldmath$\varrho$}_1$ and $\mbox{\boldmath$\varrho$}_2$
gives here $J_1(\beta \varrho_1 \varrho_2)$ and subsequent evaluation
of integrals is similar to those for (\ref{86}). We find
\begin{equation}
\displaystyle{I_4^{(2)}=-\frac{1}{4\pi}\int_{0}^{\infty}dt_1
\int_{0}^{\infty}dt_2
\exp \left(-i(t_1+t_2+T) \right) \left[N^2(t_1, t_2)-\frac{1}{(t_1+t_2+T)^2}
\right]}.
\label{88}\end{equation}
The contribution into the integral with $N^2(t_1,t_2)$ gives a domain
\newline $t_1,t_2 \sim 1/\nu_0 \ll 1$. Since $T \ll 1$ as well, we can put
an exponent in this integral equal to 1. So the integral is
\begin{equation}
\displaystyle{\int_{0}^{\infty}dt_1\int_{0}^{\infty}dt_2 N^2(t_1, t_2)
=\int_{0}^{\infty}dx\int_{0}^{\infty}dy \frac{1}{(axy+b(x+y)+a)^2}
=2 \ln \frac{b}{a} =2 \ln \coth \nu T},
\label{89}\end{equation}
where $x=\nu t_1,~y=\nu t_2,~a=\sinh \nu T,~b=\cosh \nu T$. The second
term in (\ref{88}) is calculated as (see (\ref{87}))
\begin{equation}
\begin{array}{ll}
\displaystyle{\int_{0}^{\infty}dt_1\int_{0}^{\infty}dt_2
\exp \left(-i(t_1+t_2+T) \right) \frac{1}{(t_1+t_2+T)^2} \simeq
\int_{0}^{\infty}dt {\rm e}^{-it}\frac{t}{(t+T)^2}}\\
\displaystyle{\simeq \int_{0}^{\infty}dt {\rm e}^{-it}\frac{t}{(t+T)} -1
\simeq -C - \ln T -1 + i\frac{\pi}{2}},
\end{array}
\label{90}\end{equation}
where $t=t_1+t_2$. Putting together (\ref{89}) and (\ref{90}) we
have
\begin{equation}
\displaystyle{I_4^{(2)}=2 \ln \tanh \nu T -C - \ln T -1 + i\frac{\pi}{2}}.
\label{91}\end{equation}
For computation of $I_1^{(2)}=I_3^{(2)}$ we will use Eq.(\ref{57})
\begin{equation}
\displaystyle{I_1^{(2)}=\frac{\nu^2}{4\pi} \int_{0}^{\infty} dt_1
\int_{0}^{T} dt_2 \exp \left(-i(t_1+t_2) \right)
\left[\frac{1}{\left(\nu t_1 +\nu t_2 \right)^2}-
\frac{1}{\left(\sinh \nu t_2 + \nu t_1 \cosh \nu t_2 \right)^2}\right]}
\label{92}\end{equation}
Integrating by parts over $t_1$ with regard for $\exp(-it_2) \simeq 1$
we have
\begin{equation}
\begin{array}{ll}
\displaystyle{I_1^{(2)} \simeq \frac{\nu}{4\pi}\int_{0}^{T}dt_2
\left[\frac{1}{\nu t_2}-\frac{1}{\cosh \nu t_2 \sinh \nu t_2} \right]}\\
\displaystyle{+\frac{\nu}{4\pi i}\int_{0}^{\infty}dt_1\int_{0}^{T}dt_2
\exp (-it_1) \left[\frac{1}{\nu(t_1+t_2)}-
\frac{1}{\cosh \nu t_2(\sinh \nu t_2+\nu t_1\cosh \nu t_2)} \right]}.
\end{array}
\label{93}\end{equation}
The second of these integrals is proportional to $T$ and can be neglected.
The first integral gives
\begin{equation}
\displaystyle{I_1^{(2)}=I_3^{(2)}=\frac{1}{4\pi}\ln
\frac{\nu T}{\tanh \nu T}}.
\label{94}\end{equation}
For a calculation of $I_2^{(2)}$ we use formulae (\ref{54}) and (\ref{60})
\begin{equation}
\begin{array}{ll}
\displaystyle{I_2^{(2)} \simeq \frac{\nu^2}{4\pi}\int_{0}^{T}dt(t-T)
\left[\frac{1}{\sinh^2\nu t}-\frac{1}{(\nu t_2)^2} \right]}\\
\displaystyle{=\frac{\nu}{4\pi} \int_{0}^{T}dt\left[\coth \nu t-
\frac{1}{\nu t}\right]=\frac{1}{4\pi}\ln \frac{\sinh \nu T}{\nu T}}.
\end{array}
\label{95}\end{equation}
Combining all the contributions of four domains we obtain finally
\begin{equation}
\displaystyle{\frac{dw}{d\omega}=\sum_{n=1}^{4} \frac{4\alpha}{\omega}
{\rm Re} \left(r_1 I_n^{(1)}+r_2 I_n^{(2)} \right)=
\frac{\alpha}{\pi \omega}{\rm Re}\left[r_1+
\left(\ln (\nu \sinh \nu T)-1-C\right)r_2\right]}.
\label{96}\end{equation}
In the used units ($T=al/2$) the formation length (\ref{1}) is
(see also (\ref{5}) and (\ref{14}))
\begin{equation}
\displaystyle{t_c=\frac{al_c}{2}
=\frac{1}{\zeta_c}=\varrho_c^2=\frac{1}{\nu_0+1}}.
\label{97}\end{equation}

In the case of thick target ($T \gg t_c,~\nu_0T \gg 1$) we have from
(\ref{96})
\begin{equation}
\displaystyle{\frac{dw}{d\omega} \simeq
\frac{\alpha}{\pi \omega}\left[r_1+
\left(\ln \nu_0 -1-C-\ln 2\right)r_2\right]+\frac{\alpha T}{\pi \omega}
r_2 \frac{\nu _0}{\sqrt{2}},~\frac{\alpha T}{\pi \omega}
=\frac{\alpha al}{2\pi \omega}=\frac{\alpha}{2\pi \gamma^2}
\frac{\varepsilon}{\varepsilon'}l}.
\label{98}\end{equation}
This formula gives the probability of radiation at $\nu_0 \gg 1$
(see (\ref{25}), (\ref{39})) where the contribution of boundary
photons (\ref{64}) is included.

In the case of thin target $\nu_0T \ll 1$ but when $\nu_0^2T \gg 1$
we have from (\ref{96}) the probability (\ref{79}) without term
$\propto 1/L_t$. So we have $(\nu_0^2T=4k)$
\begin{equation}
\displaystyle{\frac{dw}{d\omega} \simeq
\frac{\alpha}{\pi \omega}\left[r_1+
\left(\ln (\nu_0^2 T)-1-C\right)r_2\right]}.
\label{99}\end{equation}
Note, that when the value of the parameter $\nu_0$ is not very large,
the accuracy of the formulae (\ref{98}) and (\ref{99}) may be insufficient.
In this case one have to compute the next terms of the expansion,
as it was done in Sections 4 and 5 (see (\ref{64}) and (\ref{79})).
The same is true for (\ref{96}). A detailed analysis of the
probability of radiation in the targets of an intermediate thickness
will be carry out elsewhere.

\section{A qualitative behavior of the spectral intensity of radiation}
\setcounter{equation}{0}

We consider the spectral intensity of radiation for the energy of
the initial electrons when the LPM suppression of the intensity of radiation
takes place for relatively soft energies of photons:
$\omega \leq \omega_c \ll \varepsilon$:
\begin{equation}
\displaystyle{\nu_0(\omega_c)=1,\quad \omega_c=\frac{16\pi Z^2 \alpha^2}{m^2}
\gamma^2 n \ln \frac{a_{s2}}{\lambda_c}},
\label{100}\end{equation}
see Eqs.(\ref{9}), (\ref{13}), (\ref{14}) and (\ref{32a}).
This situation corresponds to
the \newline experimental conditions.

A ratio of a thickness of a target and the formation length of
radiation (\ref{1}) is an important characteristics of the process.
If we take into account the multiple scattering and the polarization
of a medium then the formation length (\ref{42a}) has the form
\begin{equation}
l_f = \frac{2\gamma^2}{\omega}\left[1+\gamma^2\vartheta_c^2 +
\left(\frac{\gamma \omega_0}{\omega}\right)^2 \right]^{-1},
\label{100a}\end{equation}
this ratio may be written as
\begin{equation}
\begin{array}{ll}
\displaystyle{\beta(\omega)=T\left(\nu_0+\kappa \right)\simeq T_c\left[
\frac{\omega}{\omega_c}+\sqrt{\frac{\omega}{\omega_c}}+
\frac{\omega_p^2}{\omega \omega_c} \right]},\\
\displaystyle{T=\frac{l \omega}{2\gamma^2},\quad
\omega_p=\omega_0 \gamma,\quad
T_c \equiv T(\omega_c)
\simeq \frac{2\pi}{\alpha}\frac{l}{L_{rad}}},
\end{array}
\label{101}\end{equation}
where we put that $\displaystyle{\nu_0
\simeq \sqrt{\frac{\omega_c}{\omega}}}$.
Below we assume that $\omega_c \gg \omega_p$ which is true under
the experimental conditions.

If $\beta(\omega_c)=2T_c \ll 1$
then at $\omega=\omega_c$ a target is thin and the Bethe-Heitler
spectrum of radiation, which is valid at $\displaystyle{\omega \gg \omega_c
\left(\frac{dI(\omega)}{d\omega}={\rm const}\right)}$
will be also valid at $\omega \leq
\omega_c$ in accordance with Eqs.(\ref{72}) and (\ref{72a}) since
$4k=\nu_0^2T=T_c \ll 1$. This behavior of the
spectral curve will continue with $\omega$
decrease until photon energies where a contribution of the
transition radiation become essential.

If $\beta(\omega_c) \gg 1~(T_c \gg 1)$ then at $\omega \geq \omega_c$
a target is thick and one has the LPM suppression for $\omega \leq \omega_c$.
There are two opportunities depending on the minimal value of the parameter
$\beta$.
\begin{equation}
\displaystyle{\beta_m \simeq \frac{3}{2}T_c
\sqrt{\frac{\omega_1}{\omega_c}},\quad
\omega_1=\omega_p\left(\frac{4\omega_p}{\omega_c}\right)^{1/3},\quad
\beta_m \simeq 2T_c\left(\frac{\omega_p}{\omega_c} \right)^{2/3}}.
\label{102}\end{equation}
If $\beta_m \ll 1$ then for photon energies $\omega > \omega_1$ it will
be $\omega_2$ such that
\begin{equation}
\beta(\omega_2)=1,\quad \omega_2 \simeq \frac{\omega_c}{T_c^2}
\label{103}\end{equation}
and for $\omega < \omega_2$ the thickness of a target becomes smaller
than the formation length of radiation so that for $\omega \ll \omega_2$
the spectral distribution of the radiation
intensity is described by formulae of Section 5. Under these conditions for
$4k=\nu_0^2T=T_c \gg 1$ the spectral curve has a plateau 
\begin{equation}
\displaystyle{\frac{dI}{d\omega}=\frac{2\alpha J}{\pi}={\rm const}}
\label{103a}\end{equation}
in accordance with (\ref{77}).
Such behavior of the spectral curve (first discussed in \cite{18a})
will continue until photon energies
where one has to take into account the polarization of a medium
and connected with it a contribution of the transition radiation.

At $\beta_m \gg 1$ a target remains thick for all photon energies
and radiation is described by formulae of Sections 2 and 3.
In this case at $\omega \ll \omega_c~(\nu_0 \gg 1)$ and
$\displaystyle{\omega \gg \left(\omega_p/\omega_c \right)^{1/3}\omega_p~
(\nu_0 \gg \kappa)}$ the spectral intensity of radiation formed inside
a target is given by Exp.(\ref{35}) and (\ref{39}) and the contribution
of the boundary photons is given by (\ref{64}).

Since a contribution into the spectral intensity of radiation from a passage
of an electron
inside of a target ($\propto T$) is diminishing and a contribution of
the boundary photons is increasing with $\omega$ decrease, the spectral curve
has a minimum at $\omega=\omega_m$. The value of $\omega_m$ may
be estimated from equation
\begin{equation}
\begin{array}{ll}
\displaystyle{\frac{d}{d\omega}\left(\frac{\nu_0 T}{\sqrt{2}}+\ln \nu_0
+ \frac{\pi^2 \sqrt{2}}{24}\frac{\kappa}{\nu_0} \right)=0,\quad
\frac{\nu_0T}{\sqrt{2}} \simeq 1+\frac{\pi^2 (\kappa-1)}{4\sqrt{2}\nu_0}},\\
\displaystyle{T_c \simeq \left(\frac{2 \omega_c}{\omega_p} \right)^{1/2}
\sqrt{x}+\frac{\pi^2}{4}x^2,\quad x=\frac{\omega_p}{\omega}}.
\end{array}
\label{104}\end{equation}
When a value of $T_c$ is high enough, the solution of Eq.(\ref{104})
doesn't satisfy the condition $\nu_0 \gg \kappa$ and in this case
the equation (\ref{104}) ceases to be valid.  For determination of
$\omega_m$ in this case we use the behavior of the spectral intensity of
radiation at $\kappa \gg \nu_0$. In this case a contribution into
radiation from inside passage of a target is described by (\ref{47})
whilst the radiation of the boundary photons reduces to the transition
radiation and its contribution is given by (\ref{61}). Leaving the dominant
terms ($\nu_0^2T$ is $\omega$-independent) we have
\begin{equation}
\displaystyle{\frac{d}{d\omega}\left(\frac{\nu_0^2T}{3\kappa}
+\ln \kappa \right)=0,\quad \frac{\nu_0^2T}{3\kappa} = 1,\quad
\kappa_m=\frac{T_c}{3},\quad \omega_m \simeq \sqrt{\frac{3}{T_c}}\omega_p}.
\label{105}\end{equation}
Since the value $\pi^2/12 \simeq 0.8$ is of the order of unity,
the solution of (\ref{104}) at
\newline $\kappa_m \gg \nu_0$ differs only slightly
from $\omega_m$. Because of this, if the condition
\newline $\displaystyle{2T_c(\omega_p/\omega_c)^{2/3} \gg 1}$
is fulfilled, the position of
the minimum is defined by Eq.(\ref{104}).

\section{Discussion and conclusions}
\setcounter{equation}{0}

Now we consider the experimental data \cite{12}-\cite{14} from a point
of view of the above analysis. It is shown that the mechanism of radiation
depends strongly on the thickness of a target. So, we start with an estimate
of thickness of used targets in terms of the formation length of radiation.
From Eq.(\ref{101})  we have that
\[
T_c = \frac{2\pi l}{\alpha L_{rad}} \geq 20\quad {\rm at}\quad
\frac{l}{L_{rad}} \geq  2\%.
\]
The minimum value of the ratio of a thickness of a target to the formation
length of radiation is given by Eq.(\ref{102})
($\displaystyle{\beta_m \simeq 2T_c (\omega_p/\omega_c)^{2/3}}$). For defined value of $T_c$ this ratio is
least of all for the heavy elements. Indeed, the value of $\omega_p=\omega_0
\gamma$ depends weakly on nucleus charge $Z$ ($\omega_0=30 \div 80~eV$),
while $\displaystyle{\omega_c
=\frac{4\pi \gamma^2}{\alpha L_{rad}} \propto Z^2}$.
Furthermore, the ratio $\omega_p/\omega_c$  decreases with energy increase.
Thus, among all targets with thickness $l \geq 2\% L_{rad}$
the minimal value of $\beta_m$ is attained for the heavy elements
(W, Au, U) at the initial electron energy $\varepsilon=25~GeV$.
In this case one has $\omega_c \simeq 250~MeV,
\quad \omega_p \simeq 4~MeV,\quad \beta_m \geq 2.5$. Since the parameter
$T_c$ is energy independent and the ratio $\omega_p/\omega_c \propto
1/\varepsilon$, the minimal value $\beta_m \geq 5$ is attained
at the initial electron energy $\varepsilon=8~GeV$ for
all targets with thickness $l \geq 2\% L_{rad}$ which can be considered
as thick targets.

As an example of obtained results we calculated
the spectrum of the intensity of radiation in the tungsten target
with thickness $l = 2\% L_{rad}$
at the initial electron energy $\varepsilon=8~GeV$ and $\varepsilon=25~GeV$.
The characteristic parameters of the radiation process for this case
are given in the Table. We calculated the main (Migdal) term
(Eq.(\ref{25})), the correction term (Eqs.(\ref{30}),(\ref{36}))
taking into account
an influence of the polarization of a medium according to (Eq.(\ref{45})),
as well as Coulomb corrections entering the parameters $\nu_0$
(Eq.(\ref{9a})) and $L(\varrho_c)$ (Eq.(\ref{32})). The contribution
of an inelastic scattering of a
projectile on atomic electrons (quite small for the heavy elements)
is not included although this could be done using Eq.(\ref{40a}).
We calculated also the contribution of the boundary photons Eq.(\ref{62}).
Here in the soft part of the spectrum $\omega < \omega_d
(\omega_d \simeq 2~MeV$ for
$\varepsilon=25~GeV$) the transition radiation term (\ref{61})
dominates in (\ref{62}), whilst in the harder part of the boundary photon
spectrum $\omega > \omega_d$  the terms depending on both
the multiple scattering and the polarization of a medium (\ref{59})
and (\ref{60}) give the main contribution; for $\varepsilon=8~GeV$ we
have $\omega_d \simeq 700~KeV$. It is seen that we have for
the boundary photons spectrum a smooth curve which eliminate difficulties
mentioned in \cite{14}.
All these results presented separately in Fig.2 as well as
their sum (curve 5). Note, that for energy $\varepsilon=25~GeV$ 
in the region of the minimum of the spectral curve 5 where the ratio 
of the target thickness to the formation length is minimal ($\beta_m 
\simeq 2.7$, see Table) it may be that the target is not thick enough to
use the formulae for a thick target. 
For a comparison with experiment we extract some
data from Fig.7 of \cite{14}. The theoretical curve gives the spectral
distribution of the intensity of radiation (in units $2\alpha/\pi$) without
adjusting parameters. Data from \cite{14} were recalculated according
with procedure given in it. One can see that agreement between the
experiment and theory is rather satisfactory but far from being perfect.
However, one has to take into account that the theory of LPM effect in
all previous papers had the logarithmic accuracy and did not contain
Coulomb corrections. These shortcomings did not permit to pass to the
Bethe-Heitler formula with acceptable accuracy and led to
some difficulties in data processing. Both these shortcomings are overcome
in the present paper. So, in our opinion, it is quite desirable
to handle the experimental data using the formulae of this paper.

The measurements in \cite{14} were made also using gold target
with thickness $l = 0.7\% L_{rad}$. For this case one has $T_c \simeq 6,\quad
\beta_m(25) = 0.7,\quad \beta_m(8) = 1.5$, so we have here a target of
an intermediate thickness (see Section 6). We want to stress once more
that for estimation of an effective thickness one have to use the formation
length with regard for the multiple scattering and the
polarization of a medium (see (\ref{42a}) and (\ref{100a})).
A detailed calculation
for this case will be published elsewhere.

\newpage
\setcounter{equation}{0}
\Alph{equation}
\appendix

\section{Appendix}

{\Large {\bf A potential $V(\mbox{\boldmath$\varrho$})$
with the Coulomb corrections}}
\vskip 0.6cm
It is well known, that for heavy elements the Coulomb correction
to the cross section
of bremsstrahlung of high energy particles (correction to the Born
approximation) is quite sizable, see e.g. Eq.(18.30) in \cite{7}.
The Coulomb correction (order of one for heavy elements)
is subtracted from the "large"
logarithm and if an accuracy of calculation goes beyond logarithmic one,
one has to take into account this correction.
For tungsten $(Z=74)$, gold $(Z=79)$ and uranium $(Z=92)$ in the case of
complete screening the relative Coulomb corrections to 
the standard Bethe-Heitler
cross section are respectively -7.5\% , -8.3\% and -10.7\%.

We consider the problem using eikonal approximation (see e.g.Appendix E
in \cite{7}). An amplitude $f({\bf q})$ and
a cross section of scattering in this
approximation have the form:
\begin{equation}
\begin{array}{ll}
\displaystyle{f({\bf q})=\frac{1}{2\pi i}\int_{}^{}d^2\varrho
\exp (-i{\bf q}\mbox{\boldmath$\varrho$}) S(\mbox{\boldmath$\varrho$}),\quad
S(\mbox{\boldmath$\varrho$})=
\exp \left(-i\chi(\mbox{\boldmath$\varrho$})\right)-1},\\
\displaystyle{\chi(\mbox{\boldmath$\varrho$})=\int_{-\infty}^{\infty}
U(\mbox{\boldmath$\varrho$},z)dz,\quad
d\sigma({\bf q})=|f({\bf q})|^2d^2q},
\end{array}
\label{a1}\end{equation}
where $(z, \mbox{\boldmath$\varrho$})$ are the
longitudinal and transverse coordinates respectively,
$U(\mbox{\boldmath$\varrho$}, z)$ is the potential.
Repeating a derivation made in Section 2 (eqs. (\ref{3})-(\ref{9})) but with
the cross section (\ref{a1}) we find for the potential
$V(\mbox{\boldmath$\varrho$})$
\begin{equation}
\begin{array}{ll}
\displaystyle{V(\mbox{\boldmath$\varrho$})=n\int_{}^{}\left(1-
\exp (i{\bf q} \mbox{\boldmath$\varrho$}) \right)|f({\bf q})|^2 d^2q}\\
\displaystyle{=n\int_{}^{}d^2x \left(S({\bf x})S^{\ast}({\bf x})-
S({\bf x}+\mbox{\boldmath$\varrho$})S^{\ast}({\bf x}) \right)}.
\end{array}
\label{a2}\end{equation}
Since the potential $V(\mbox{\boldmath$\varrho$})$ was calculated above
in the Born approximation, we can calculate here the difference of the
potentials calculated in the Born approximation
$V_B(\mbox{\boldmath$\varrho$})$ and in the eikonal
approximation $V(\mbox{\boldmath$\varrho$})$
\begin{equation}
\begin{array}{ll}
\Delta V(\mbox{\boldmath$\varrho$})=V_B(\mbox{\boldmath$\varrho$})-
V(\mbox{\boldmath$\varrho$})=n\int_{}^{}d^2x\Big\{\exp\left[
i\chi(\mbox{\boldmath$\varrho$}+{\bf x})-i\chi(x) \right]
-1 \\
+\frac{1}{2}\left[\chi(\mbox{\boldmath$\varrho$}+{\bf x})-
\chi(x) \right]^2 \Big\};\\
\displaystyle{\chi(x)=\int_{-\infty}^{\infty} dz \frac{Z\alpha}{r}
\exp \left(-\frac{r}{a_s}\right)
=2Z\alpha K_0\left(\frac{x}{a_s}\right),\quad r=\sqrt{z^2+x^2}},
\end{array}
\label{a3}\end{equation}
where $K_0(z)$ is the modified Bessel function.
Because the eikonal phase enters (\ref{a3}) only in the
combination $\chi(\mbox{\boldmath$\varrho$}+{\bf x})-
\chi(x)$ in the interesting for us region $x \sim \varrho$.
Since $\displaystyle{K_0\left(\frac{x}{a_s}\right)}$ is large only if
$x/a_s \ll 1$ it is evident that main contribution into integral
(\ref{a3}) gives the region $x \sim \varrho \ll a_s$. In this region
one has
\begin{equation}
\displaystyle{\chi(\mbox{\boldmath$\varrho$}+{\bf x})-\chi(x)=
\xi \ln \frac{x^2}{(\mbox{\boldmath$\varrho$}+{\bf x})^2},\quad
\xi=Z\alpha}.
\label{a4}\end{equation}
So, in the expression for $\Delta V(\mbox{\boldmath$\varrho$})$ enters
only one dimensional parameter $\mbox{\boldmath$\varrho$}={\bf l}\varrho$,
where ${\bf l}$ is the unit vector. After substitution of variables
${\bf x} \rightarrow \varrho {\bf x}$ we have
\begin{equation}
\displaystyle{\frac{\Delta V(\mbox{\boldmath$\varrho$})}{n}=
2\pi \varrho^2 f(\xi),\quad f(\xi)=\frac{1}{2\pi}\int_{}^{}d^2x
\left[\left(\frac{({\bf x}+{\bf l})^2}{x^2} \right)^{i\xi} - 1
+ \frac{\xi^2}{2}\ln^2 \frac{({\bf x}+{\bf l})^2}{x^2} \right]}.
\label{a5}\end{equation}
Changing variables $\displaystyle{{\bf y}=\frac{{\bf x}}{x^2}}$ and then
${\bf z} = {\bf y}+{\bf l}$ we have
\begin{equation}
\displaystyle{f(\xi)=\frac{1}{2\pi}\int_{}^{}\frac{d^2z}{{(\bf z}-{\bf l})^4}
\left[z^{2i\xi}-1+\frac{\xi^2}{2}\ln^2 z^2 \right]}.
\label{a6}\end{equation}
Integration over azimuthal angle gives
\begin{equation}
\displaystyle{\int_{0}^{2\pi}\frac{d\phi}{(z^2-2z \cos \phi + 1)^2}=
\frac{2\pi(1+z^2)}{|z^2-1|^3}}.
\label{a7}\end{equation}
Changing the variable $z^2=u$, splitting the integration interval
into two parts:
\newline $(0,1)$ and $(1,\infty)$ changing in the second interval
$v=1/u$ we obtain
\begin{equation}
\displaystyle{f(\xi)={\rm Re}\int_{0}^{1}\frac{du(1+u)}{(1-u)^3}
\left(u^{i\xi}-1+\frac{\xi^2}{2}\ln^2 u \right)}
\label{a8}\end{equation}
Integrating by parts and changing once more variable $u=e^{-y}$
we find
\begin{equation}
\displaystyle{f(\xi)={\rm Re}\int_{0}^{\infty}\frac{e^{-y}dy}{(1-e^{-y})^2}
\left[-i\xi e^{-i\xi y}+\xi^2 y \right]}.
\label{a9}\end{equation}
Integrating once more by parts and using the standard (Gauss) representation
of the Euler $\psi$-function we have finally
\begin{equation}
\displaystyle{\Delta V(\varrho)=2\pi\varrho^2 f(Z\alpha),\quad
f(\xi)=\xi^2 {\rm Re} \left[\psi(1+i\xi)-\psi(1) \right] =
\xi^2 \sum_{n=1}^{\infty}\frac{1}{n(n^2+\xi^2)}.}
\label{a10}\end{equation}
The obtained function $f(\xi)$ is the known Coulomb correction
to the Bethe-Heitler cross section of bremsstrahlung, see e.g. \cite{7},
Sections 17,18.

\setcounter{equation}{0}

\section{Appendix}

{\Large {\bf An allowance for a form factor of a nucleus}}
\vskip0.6cm
When $\varrho_c \ll R_n$ (see (\ref{14}), (\ref{15})) one cannot
consider the potential of a nucleus as a potential of a point charge.
A contribution into the multiple scattering gives a momentum transfer
$q \leq 1/R_n$. Because of the same reason the phase
${\bf q} \mbox{\boldmath$\varrho$}$ in expression (\ref{a2})
for the potential $V(\mbox{\boldmath$\varrho$})$ is small
$q\varrho \leq \varrho_c/R_n \ll 1$ and one can expand the potential.
As a result we obtain
\begin{equation}
\begin{array}{ll}
\displaystyle{V(\varrho)=\frac{n\varrho^2}{4}\int_{}^{}|{\bf q}f(q)|^2d^2q=
\frac{\varrho^2}{4}\int_{}^{}|\mbox{\boldmath$\nabla$} S({\bf x})|^2d^2x}\\
\displaystyle{=\frac{\varrho^2}{4}\int_{}^{}
\left(\mbox{\boldmath$\nabla$} \chi({\bf x})\right)^2d^2x=
\frac{\varrho^2}{4}\int_{}^{}
\left({\bf q}({\bf x})\right)^2d^2x},
\end{array}
\label{b1}\end{equation}
where ${\bf q}({\bf x})$ is the classical momentum transfer on a
straight-line trajectory with an impact parameter ${\bf x}$.
As one can see from (\ref{b1}), the mean square of the momentum transfer
is the same the in eikonal approximation, in the Born approximation
and in the classical theory. The Coulomb correction in this case vanishes.

Considering a nucleus as an uniformly charged sphere with the radius
$R_n$, we have
\begin{equation}
\begin{array}{ll}
\displaystyle{{\bf q}(\mbox{\boldmath$\varrho$})=
\frac{2\xi \mbox{\boldmath$\varrho$}}{\varrho^2}\left[
\frac{\varrho}{a_s}K_1\left(\frac{\varrho}{a_s}\right)
\vartheta\left(\varrho-R_n\right)+\varphi\left(
\frac{\varrho^2}{R_n^2} \right)
\vartheta\left(R_n-\varrho\right)\right]},\\
\displaystyle{\varphi(x)=1-\sqrt{1-x}+x \sqrt{1-x}}.
\end{array}
\label{b2}\end{equation}
Substituting the expression obtained into (\ref{b1}) we find
the potential $V(\varrho)$ under conditions considered
\begin{equation}
\begin{array}{ll}
\displaystyle{\int_{}^{}q^2(\varrho)d^2\varrho=4\pi \xi^2
\left[\frac{2}{a_s^2}\int_{R_n}^{\infty}K_1^2\left(
\frac{\varrho}{a_s} \right)\varrho d\varrho +
\int_{0}^{1}\frac{dx}{x} \varphi^2(x)\right]
=4\pi \xi^2 \bigg[\bigg(\ln \left(\frac{2a_s}{R_n} \right)^2}\\
\displaystyle{-1-2C \bigg)+\bigg(\frac{7}{2}-4 \ln 2 \bigg) \bigg]
=4\pi Z^2\alpha^2 \left[\ln \frac{a_s^2}{R_n^2}+
\frac{5}{2} -2(C +\ln 2)\right]};\\
\displaystyle{V(\varrho)=\pi \xi^2 n \varrho^2\left[\ln \left(
\frac{a_s^2}{R_n^2} \right)-0.0407 \right]}.
\end{array}
\label{b3}\end{equation}

If one uses standard representation of nuclear form factor (see e.g.
\cite{17})
\begin{equation}
\displaystyle{F(q)=\frac{1}{(1+q^2\varrho_0^2)^2},\quad
\varrho_0^2=\frac{R_n^2}{6},
\quad \varrho_N = 1.2 \cdot 10^{-13} A^{1/3} cm},
\label{b4}\end{equation}
then one obtains
\begin{equation}
\displaystyle{\int_{}^{}q^2(\varrho)d^2\varrho
=4\pi Z^2\alpha^2 \left[\ln \frac{a_s^2}{R_n^2}+\ln 6 -2 \right]
\simeq 4\pi Z^2\alpha^2 \left[\ln \frac{a_s^2}{R_n^2}-0.208 \right]}.
\label{b5}\end{equation}
Taking into account that $\displaystyle{\ln \frac{a_s^2}{R_n^2}
\simeq 20}$ we
see that the difference between different models of nucleus is less than
1 \% .

\newpage

{\bf Figure captions}

\vspace{15mm}
\begin{itemize}

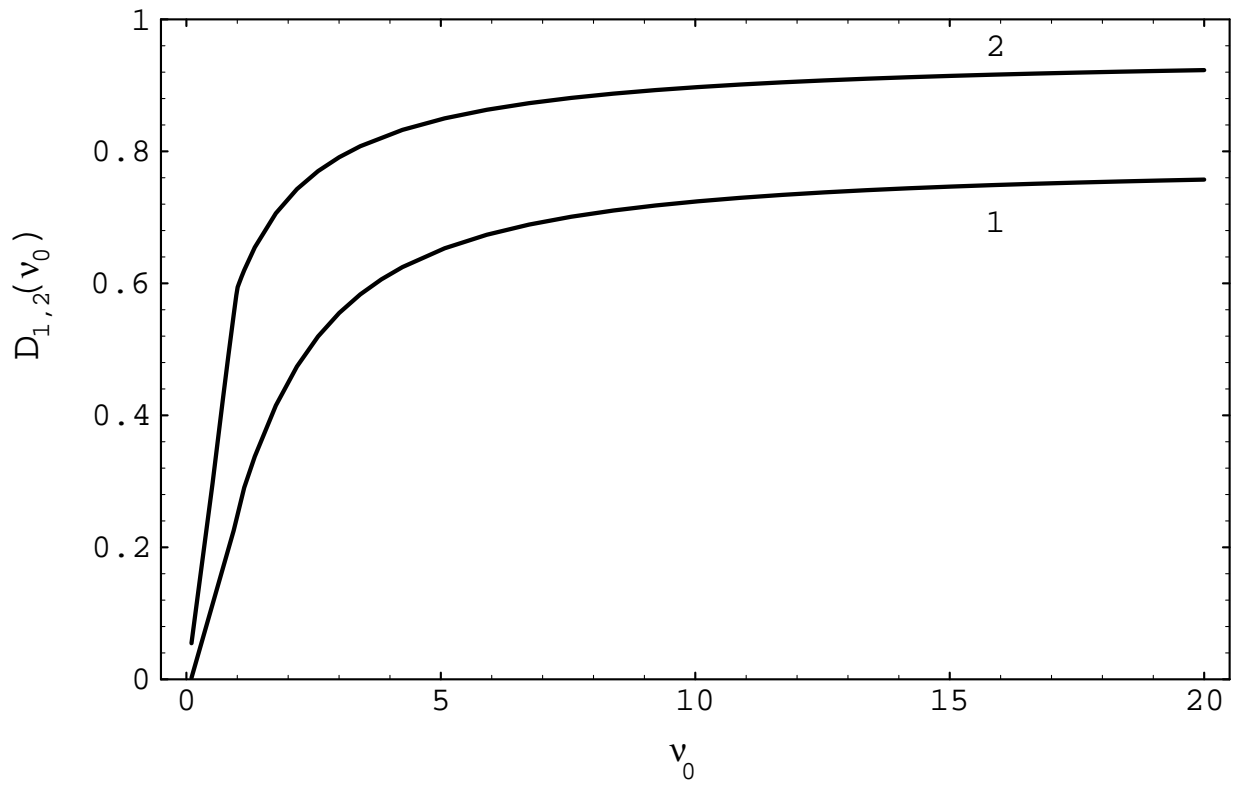
\item {\bf Fig.1}  The functions $D_{1,2}(\nu_0)$ (Eq.(\ref{36})) vs
parameter $\nu_0$.

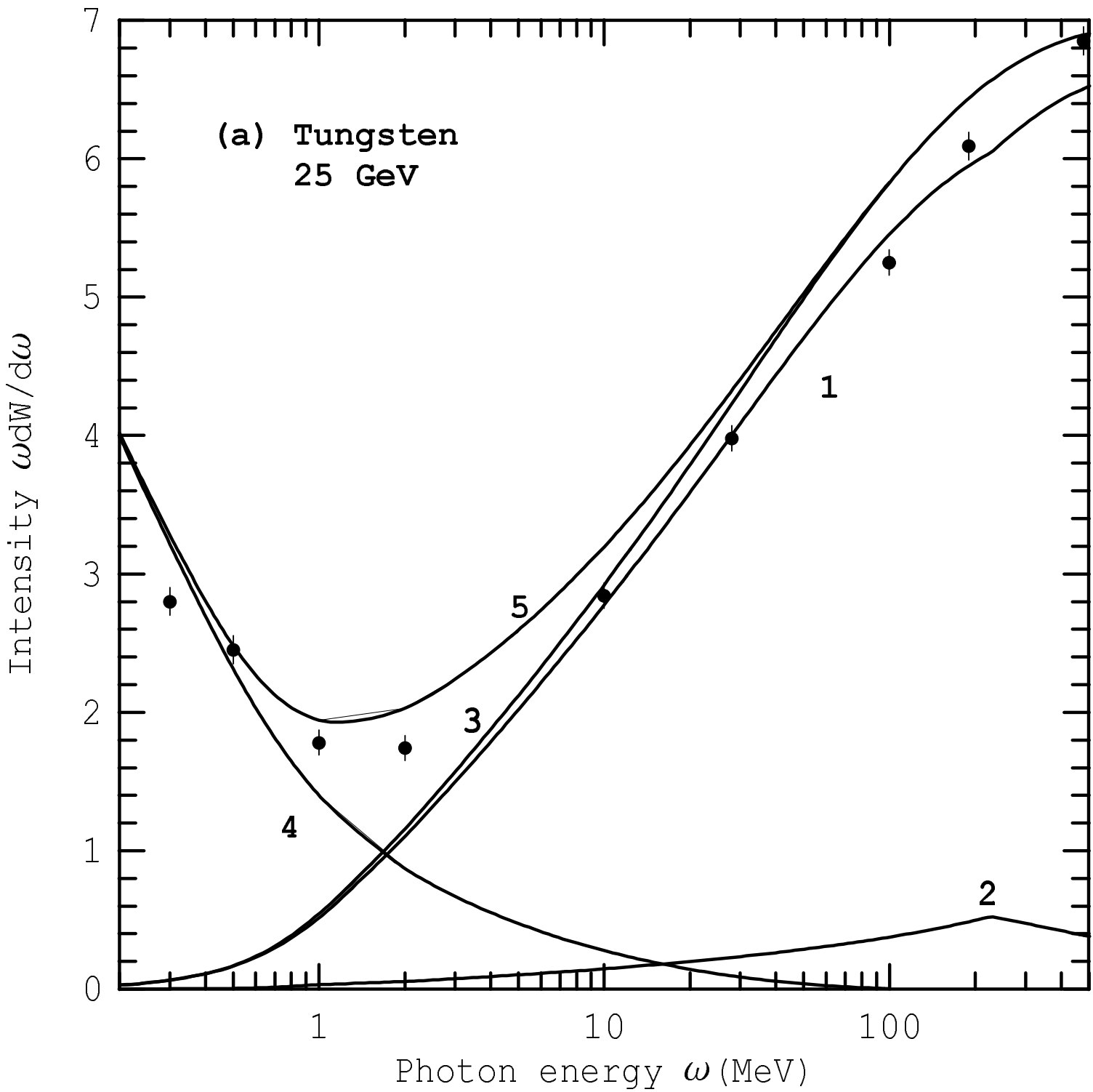
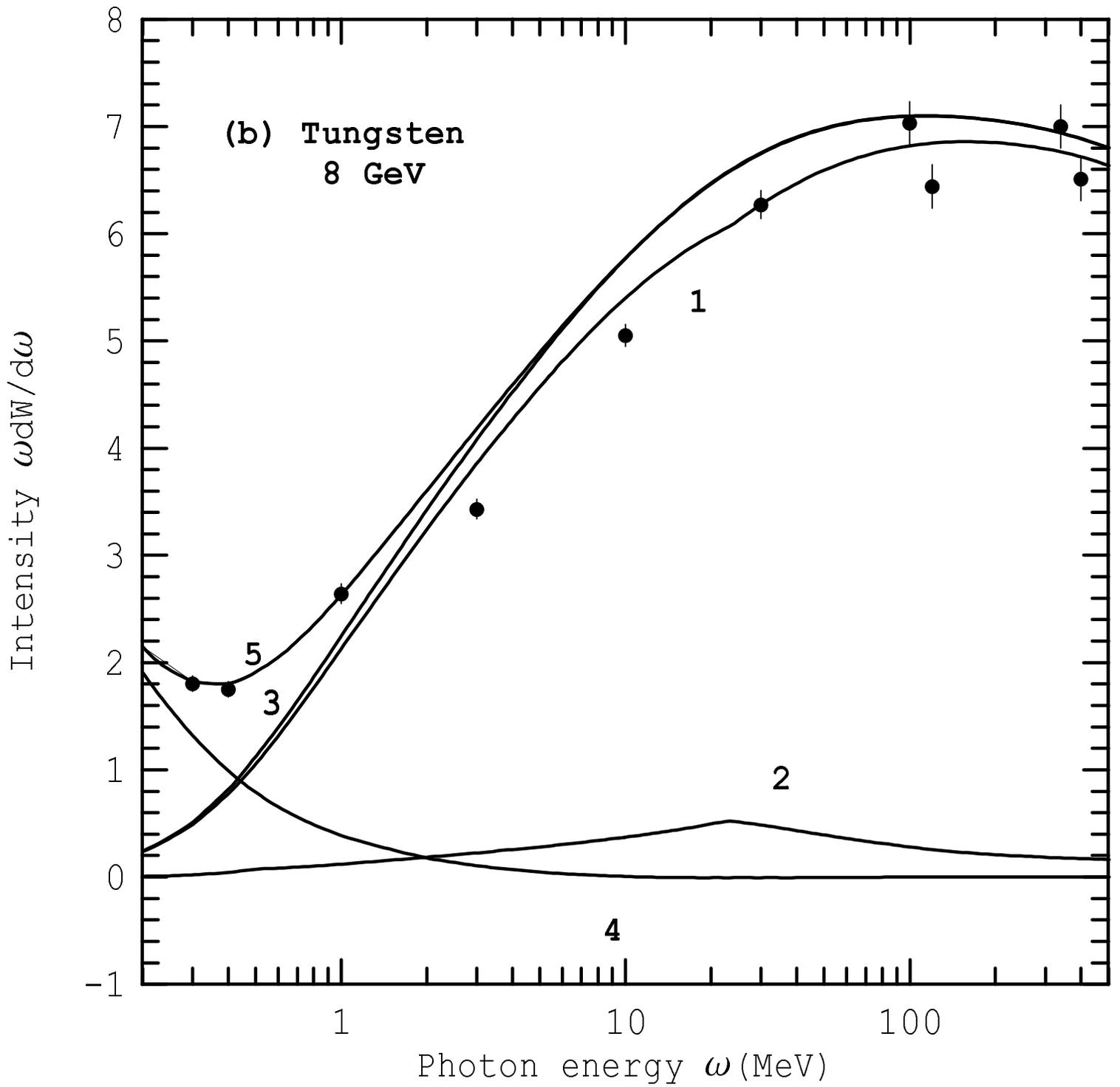
\item {\bf Fig.2} The intensity of radiation
$\displaystyle{\omega \frac{dW}{d\omega}}$
in tungsten with thickness $l=0.088~mm$ in units
$\displaystyle{\frac{2\alpha}{\pi}}$,
((a) is for the initial electrons energy $\varepsilon=25~GeV$ and (b)
is for $\varepsilon=8~GeV$).
The Coulomb corrections and the polarization of a medium are included.
\begin{itemize}
\item Curve 1 is the contribution of the main term (\ref{25});
\item curve 2 is the correction (\ref{30}), (\ref{36});
\item curve 3 is the sum of the previous contributions;
\item curve 4 is the contribution of the boundary photons (\ref{62});
\item curve 5 is the total prediction for the intensity of radiation.
\end{itemize}

\end{itemize}

\newpage

\begin{table}
\begin{center}
{\sc TABLE}~
{Characteristic parameters of the radiation process in}\\
{tungsten with the thickness $l = 2\% L_{rad}$}
\end{center}
\begin{center}
\begin{tabular}{*{7}{|c}|}
\hline
$\varepsilon~(GeV)$ & $\omega_c~(MeV)$ & $\omega_p~(MeV)$& $T_c$&
$\omega_1~(MeV)$& $\beta_m$&$\omega_m~(MeV)$ \\ \hline
25 & 228  &  3.93& 21.25&  1.6 &  2.7&  2 \\ \hline
 8 & 23.35&  1.26& 21.25&  0.76&  5.7& 0.5\\ \hline
\end{tabular}
\end{center}
\end{table}


\newpage

\begin{thebibliography}{99}
\bibitem{1} L. D. Landau and I. Ya. Pomeranchuk, Dokl.Akad.Nauk SSSR
{\bf 92} (1953) 535, 735. See in English in {\em The
Collected Paper of L. D. Landau}, Pergamon Press, 1965.
\bibitem{2} A. B. Migdal, Phys. Rev. {\bf 103} (1956) 1811.
\bibitem{3} A. B. Migdal, Sov. Phys. JETP {\bf 5} (1957) 527.
\bibitem{4} M. L. Ter-Mikaelian, {\em High Energy Electromagnetic
Processes in Condensed Media}, John
Wiley \& Sons, 1972.
\bibitem{5} E. L. Feinberg and I. Ya. Pomeranchuk, Nuovo Cimento,
Supplement to Vol. {\bf 3} (1956) 652.
\bibitem{6} V. M. Galitsky and I. I. Gurevich, Nuovo Cimento
{\bf 32} (1964) 396.
\bibitem{7} V. N. Baier, V. M. Katkov and V. S Fadin,
{\em Radiation from Relativistic Electrons} (in Russian) Atomizdat,
Moscow, 1973.
\bibitem{8} V.N.Baier, V.M.Katkov and V.M.Strakhovenko,
Sov. Phys. JETP {\bf 67} (1988) 70.
\bibitem{9} V.N.Baier, V.M.Katkov and V.M.Strakhovenko,
{\em Electromagnetic Processes at High Energies in Oriented
Single Crystals}, World Scientific Publishing Co, Singapore, 1997.
\bibitem{10} B. G. Zakharov, Pis'ma v ZhETF {\bf 63} (1996) 906.
\bibitem{11} R.Blancenbeckler and S. D. Drell, Phys.Rev.
{\bf D53} (1996) 6265.
\bibitem{12} P. L. Anthony, R. Becker-Szendy, P. E. Bosted {\em et al},
Phys. Rev. Lett. {\bf 75} (1995) 1949.
\bibitem{13} P. L. Anthony, R. Becker-Szendy, P. E. Bosted {\em et al},
Phys. Rev. Lett. {\bf 76} (1996) 3550.
\bibitem{14} P. L. Anthony, R. Becker-Szendy, P. E. Bosted {\em et al},
Preprint SLAC-PUB-4713, February 1997 (Submitted to Phys.Rev.D)
\bibitem{15} M. P. Perl in {\em Proc. 1994 Les Rencontres de Physique
de la Vallee D'Aoste}, (Editions Frontieres, Gif-sur-Yvette, France, 1994).
Ed.M. Greco,p. 567.
\bibitem{16} V. B. Berestetskii, E. M. Lifshitz and L. P. Pitaevskii,
{\em Quantum Electrodynamics} Pergamon Press, Oxford, 1982.
\bibitem{17} Y.-S. Tsai, Rev. Mod. Phys. {\bf 46} (1974), 815.
\bibitem{18} V.N.Baier, V.M.Katkov and V.M.Strakhovenko,
Nucl. Phys. {\bf B328} (1989) 387.
\bibitem{16a} R.Blankenbecler
Phys.Rev {\bf D55} (1997) 190.
\bibitem{17a} I. I. Gol'dman
Sov. Phys. JETP {\bf 11} (1960) 1341.
\bibitem{18a} F. F. Ternovskii
Sov. Phys. JETP {\bf 12} (1960) 123.
\bibitem{18b}V. E. Pafomov
Sov. Phys. JETP {\bf 20} (1965) 253.
\bibitem{18c} B. G. Zakharov
Pis'ma v ZhETF {\bf 64 } (1996) 737.
\bibitem{20} V. N. Baier, V. S. Fadin, V. A. Khoze and E.A. Kuraev
Phys. Rep. {\bf 78} (1981) 293.

\end{thebibliography}
\end{document}